\documentclass{article}

\usepackage{PRIMEarxiv}

\usepackage[utf8]{inputenc} 
\usepackage[T1]{fontenc}    
\usepackage{hyperref}       
\usepackage{url}            
\usepackage{booktabs}       
\usepackage{amsfonts}       
\usepackage{amsmath}
\usepackage{nicefrac}       
\usepackage{microtype}      
\usepackage{lipsum}
\usepackage{authblk}
\usepackage{fancyhdr}       
\usepackage{graphicx}       
\usepackage{appendix}
\usepackage{enumitem}
\usepackage{titlesec}
\usepackage{tabularx}
\usepackage{makecell}
\usepackage{textcomp}
\usepackage{float}
\usepackage{subcaption}   
\usepackage{etoolbox} 
\graphicspath{{media/}}

\newsavebox{\mybox}

\title{UWB RADAR-BASED HEART RATE MONITORING:\newline A TRANSFER LEARNING APPROACH
}

\author[1,$\dagger$]{\mbox{Elzbieta Gruzewska}}
\author[1]{\mbox{Pooja Rao}}
\author[1]{\mbox{Sebastien Baur}}
\author[1,2,*]{\mbox{Matthew Baugh}}
\author[1]{\mbox{Mathias M.J. Bellaiche}}
\author[1]{\mbox{Sharanya Srinivas}}
\author[1]{\mbox{Octavio Ponce}}
\author[1]{\mbox{Matthew Thompson}}
\author[1]{\mbox{Pramod Rudrapatna}}
\author[1]{\mbox{Michael A. Sanchez}}
\author[1]{\mbox{Lawrence Z. Cai}}
\author[3,4]{\mbox{Timothy JA Chico}}
\author[3,4]{\mbox{Robert F. Storey}}
\author[3,4]{\mbox{Emily Maz}}
\author[1]{\mbox{Umesh Telang}}
\author[1]{\mbox{Shravya Shetty}}
\author[1]{\mbox{Mayank Daswani}}

\affil[1]{Google}
\affil[2]{Imperial College London}
\affil[3]{NIHR Sheffield Biomedical Research Centre, Sheffield Teaching Hospitals NHS Foundation Trust, Sheffield, United Kingdom}
\affil[4]{Division of Clinical Medicine, University of Sheffield, Sheffield, United Kingdom}

\affil[*]{Work performed while interning at Google Research}
\affil[$\dagger$]{Corresponding author: \texttt{epawelczyk@google.com}}

\begin{document}
\maketitle

%
%
\begin{abstract}
    Radar technology presents untapped potential for continuous, contactless, and passive heart rate monitoring via consumer electronics like mobile phones. However the variety of available radar systems and lack of standardization means that a large new paired dataset collection is required for each radar system. 
    This study demonstrates transfer learning between frequency-modulated continuous wave (FMCW) and impulse-radio ultra-wideband (IR-UWB) radar systems, both increasingly integrated into consumer devices. FMCW radar utilizes a continuous chirp, while IR-UWB radar employs short pulses. Our mm-wave FMCW radar operated at 60 GHz with a 5.5 GHz bandwidth (2.7 cm resolution, 3 receiving antennas [Rx]), and our IR-UWB radar at 8 GHz with a 500 MHz bandwidth (30 cm resolution, 2 Rx).
    Using a novel 2D~+~1D ResNet architecture we achieved a mean absolute error (MAE) of 0.85 bpm and a mean absolute percentage error (MAPE) of 1.42\% for heart rate monitoring with FMCW radar (N=119 participants, an average of 8 hours per participant). This model maintained performance (under 5 MAE/10\% MAPE) across various body positions and heart rate ranges, with a 98.9\% recall. We then fine-tuned a variant of this model, trained on single-antenna and single-range bin FMCW data, using a small (N=376, avg 6 minutes per participant) IR-UWB dataset. This transfer learning approach yielded a model with MAE 4.1 bpm and MAPE 6.3\% (97.5\% recall), a 25\% MAE reduction over the IR-UWB baseline.
    This demonstration of transfer learning between radar systems for heart rate monitoring has the potential to accelerate its introduction into existing consumer devices. 
\end{abstract}

%
%
\section{Introduction}
    Heart rate (HR) is a critical physiological parameter in healthcare, offering fundamental insight into an individual's cardiovascular status and physiological responses across various health conditions \cite{ref_68, ref_69, ref_70}. Both elevated and depressed heart rates are well-established indicators of the body's reaction to perturbations stemming from infectious diseases, endocrine dysfunctions, mood disorders, and a spectrum of cardiovascular conditions \cite{ref_71, ref_72, ref_73, ref_74, ref_75, ref_76}. Heart rate monitoring is also valuable in a wide array of general health and wellness applications, including exercise and fitness tracking, hydration assessment, and sleep analysis \cite{ref_77, ref_78, ref_79}.
    
    Wearable devices such as smartwatches and rings offer continuous and precise pulse monitoring, but their adoption lags far behind smartphones \cite{ref_55, ref_80}. Even among individuals who own wearables, only 43.3\% report wearing them all the time (day and night) \cite{ref_55}, likely because of inconvenience or the need for frequent charging. Technology that enables contactless vital sign monitoring via ubiquitous devices such as smartphones would extend the benefits of ambient HR monitoring to a larger population.

    Radar technology presents a promising route to integrating contactless and continuous vital sign monitoring into consumer electronics because it preserves privacy, is agnostic to skin tone, and penetrates clothes and blankets \cite{ref_82}. The radar systems that have shown the most promise for vital sign measurement from consumer devices include millimeter wave frequency-modulated continuous wave (mm-wave FMCW)\cite{ref_1, ref_2, ref_3, ref_7} and impulse-radio ultra-wideband (IR-UWB)\cite{ref_4, ref_5, ref_6} radar systems. mm-wave FMCW radar has been used in consumer electronics, including some smartphones and Google Nest Hub devices. UWB technology is growing in popularity \cite{ref_56, ref_67}, and the UWB chip is present in an expanding range of consumer electronics such as smartphones \cite{ref_39}, but its use so far has been limited to non-radar applications such as localization and tracking (including common devices such as AirTags or SmartTags) \cite{ref_66}, vehicle unlock features \cite{ref_39}, or data transfer \cite{ref_39}. As we show in this paper, utilizing the radar capability of the same UWB chip can unlock contactless vital sign measurement via the smartphone.
    
    Radar measurement of vital signs operates by detecting body micromotion, most frequently that of the chest wall. However, the amplitude of chest wall movement caused by cardiac pulsation is only 1/20th of that caused by respiratory movements \cite{ref_47}. Therefore, one of the main challenges in attempting to measure HR over long periods of time in freely moving individuals is to differentiate cardiac from respiratory and other body movements and environmental noise \cite{ref_47}. As described in the Related Work section, various researchers have proposed signal processing methods \cite{ref_16, ref_25, ref_46, ref_50, ref_61, ref_63} either alone or in combination with machine learning \cite{ref_7, ref_20, ref_21, ref_22, ref_23, ref_24, ref_25, ref_26, ref_27} to measure HR from radar signals.
    
    The accuracy of machine learning models depends on the size and quality of the training dataset, and the collection of a paired dataset for each new radar configuration is time-consuming and expensive. For example, establishing the state-of-the-art (SOTA) for FMCW radar HR \cite{ref_7} required a dataset with 980 cumulative hours of 119 participants, with the radar device operating concurrently with a reliable source of heart rate ground truth (ECG and/or PPG).
    
    We therefore propose a transfer learning technique to effectively leverage models trained on mm-wave FMCW for IR-UWB. Those two radar systems operate using completely different physical principles: mm-wave FMCW transmits a continuous sinusoidal wave whose frequency increases linearly with time, while IR-UWB transmits very short pulses with duration on the order of a few hundred picoseconds to a few nanoseconds. Moreover, these two systems usually have widely different range resolution due to the different bandwidths on which they operate. In the case of radars used here, IR-UWB radar has 10x lower range resolution compared to mm-wave FMCW radar. By ‘transferring’ features learned using one radar configuration to another, we can substantially accelerate the development of radar HR technology.
    
    This work makes the following key contributions:
    \begin{itemize}
    \item We introduce a novel deep learning framework designed to extract spatio-temporal features from radar signals. Building on previous research \cite{ref_7}, this approach advances SOTA performance in ambient heart rate measurement for FMCW radar, achieving a Mean Absolute Error (MAE) of 0.85 bpm (halving the previous SOTA on this dataset of 1.69 bpm) and a recall rate of 98.9\% (previous SOTA: 88.53\%) \cite{ref_7}. 
    \item We demonstrate transfer learning between the FMCW and IR-UWB radar modalities, leveraging a large mm-wave FMCW dataset (980 hours) to increase the accuracy of HR measurement on IR-UWB radar using a much smaller dataset (37.3 hours).
    \item This transfer learning approach allowed us to achieve clinically acceptable accuracy \cite{ref_58, ref_59, ref_60} for heart rate monitoring using a IR-UWB radar configuration similar to that in current mobile phones. The ability to perform contactless heart rate measurement via these readily available personal devices holds the potential to extend cardiovascular health monitoring to millions of users worldwide.
    \end{itemize}

%
%
\section{Related Work}
    \subsection{Heart Rate Sensing with Radar Technology}
        Over the past decade, researchers have demonstrated the feasibility of heart rate measurement, using both mm-wave Frequency Modulated Continuous Wave (FMCW) \cite{ref_1, ref_2, ref_3} radar, and pulsed Ultra-Wide Band (UWB) \cite{ref_4, ref_5, ref_6} radar. These initial studies were limited by short acquisition durations and small sample sizes, and were performed under tightly controlled laboratory conditions. For mm-wave FMCW radar, recent studies with a larger number of participants demonstrate that heart rate \cite{ref_7} and heart rate variability (HRV) \cite{ref_8} can be accurately monitored overnight, establishing the state of the art for heart rate monitoring during sleep.
    
        The potential use of radar to measure more advanced and complex cardiovascular signs such as carotid pulse waveform \cite{ref_9}, blood pressure \cite{ref_10}, and to reconstruct ECG segments \cite{ref_11, ref_12} supports the need for more research into the possibility of obtaining these important measurements using consumer-grade radar technology.
        
    \subsection{Deep neural network architectures for heart rate estimation}
        A variety of deep neural network architectures have been proposed for heart rate estimation from radar signals \cite{ref_13}. In some cases, deep learning is used in preprocessing, followed by traditional signal processing algorithms for the final heart rate computation. Examples include the use of convolutional neural networks (CNNs) to improve signal-to-noise ratio \cite{ref_14, ref_15}, select heart rate segments \cite{ref_16}, assign weights to range bins \cite{ref_17}, and the use of a variational autoencoder-decoder \cite{ref_18} to remove motion interference. Additionally, long short-term memory networks (LSTMs) have been used to discriminate between heartbeats \cite{ref_19} and other signals.
    
        More commonly, deep learning methods are applied to estimate heart rate or related signals after the data has been cleaned and preprocessed. A number of studies use CNNs, ranging from 1D CNNs \cite{ref_20, ref_21}, through off-the-shelf CNN architectures pre-trained using ImageNet \cite{ref_22, ref_23}, to partly or fully custom CNN architectures \cite{ref_7, ref_24, ref_25, ref_26, ref_27}. The choice of CNN architecture and customization method varies based on the primary goal, such as efficient real-time or mobile deployment \cite{ref_7, ref_24, ref_27}, or focusing on the extraction of features such as frequency \cite{ref_25} or a time-frequency representation \cite{ref_26}.
        
        Beyond CNNs, several studies have proposed the use of LSTMs to capture temporal dependencies, either alone \cite{ref_28, ref_29, ref_30} or in combination with a CNN \cite{ref_31, ref_32, ref_33, ref_34}. Recently, researchers have also used transformers \cite{ref_35, ref_36} or a hybrid CNN-transformer architecture \cite{ref_37}, leveraging attention-based processing to model the complex spatial-temporal relationships in radar signals for HR estimation. As shown by Tran et al. \cite{ref_38} for video analysis, CNN architectures, specifically ResNet variations, can also be used to extract spatio-temporal features through hierarchical feature learning—a concept we utilize in our proposed method.
        
    \subsection{Radar and Transfer Learning}
        In addition to fundamental differences in the type of radar (mm-wave FMCW versus IR-UWB), the studies above show variations in configuration such as the number and orientation of transmitter antennas, bandwidth, and range resolution. As the field graduates from specialized radar systems to UWB technology available in consumer electronics \cite{ref_39}, constraints such as power consumption, device space, orientation requirements, and manufacturing costs reduce flexibility compared to research environments. It is therefore important to study how properties learned in one radar configuration transfer to a new setup. Effective transfer would reduce the need for new training datasets, accelerate algorithm deployment, and improve accuracy. 
    
        Previous studies of transfer learning in radar have focused on gesture \cite{ref_40} or activity recognition \cite{ref_41, ref_42}, where researchers used CNNs pre-trained with a large image or video dataset and performed fine-tuning using a smaller task-specific radar dataset. Related ‘domain adaptation’ approaches to mitigate the lack of dataset availability involve the use of large Motion Capture databases to create simulated radar data \cite{ref_43, ref_44}, which are then used to train activity recognition models.
        
        Features extracted from ECG datasets have been used to classify arrhythmia \cite{ref_45} and improve the accuracy of heart rate detection \cite{ref_46} in mm-wave FMCW radar datasets, indicating that cardiac signals are transferable across domains.
        
        However, most of the above studies address transfer learning within different configurations of the same radar type, with a single study exploring transfer learning from IR-UWB to mm-wave FMCW radar in the context of activity recognition \cite{ref_40}. The use of transfer learning in the context of vital sign measurement has not yet been explored.
        
        \subsection{Separating Respiratory Motion from Cardiac Motion}
        While machine learning methods are increasingly used \cite{ref_13} in radar-based heart rate measurement, advances in signal processing can still provide highly accurate results \cite{ref_8}.
    
        As the amplitude of chest wall movements due to cardiac motion is approximately 1/20th of that due to respiration \cite{ref_47}, there is a need for novel methods to separate the two, in order to increase the signal-to-noise ratio for downstream ML algorithms. Band-pass filters \cite{ref_3} can filter out the dominant respiratory rate frequency, but leave higher-order harmonics that can still overshadow the HR fundamental frequency. Other methods include using the time domain to separate breathing from respiratory signals \cite{ref_16, ref_48, ref_49}, the use of adaptive filtering \cite{ref_46, ref_50}, and the use of higher-order harmonics \cite{ref_8, ref_51}. 
        
        Based on the method proposed by Srinivas et al. \cite{ref_50}, we use a filter that is derived adaptively from the respiratory rate per sample to facilitate heart rate signal extraction.

%
%
\section{Methods}
\label{sec:methods}
    \subsection{Data collection}
        \subsubsection{mm-wave FMCW radar Dataset}
        This dataset was described in Luzhou Xu et.al \cite{ref_7}. In short, a second-generation Soli radar chip integrated into Google Nest Hub devices was used to collect mm-wave FMCW data; this chip operates at 60 GHz with one transmitting and three receiving antennas. It transmits frequency-modulated continuous wave (FMCW) waveforms, also known as chirps, sweeping the frequency from 58 GHz to 63.5 GHz, for a bandwidth of 5.5 GHz. Each chirp contained 256 samples, and was repeated 20 times at a rate of 30 Hz within a burst.
        
        The data was collected in a room with a bed used by the participant to sleep overnight and the radar device on the bedside table. Each participant wore an ECG lead II sensor for tracking a ground truth HR. The recordings were performed overnight while the participant was sleeping and contained around 7-9h of recording per session. While 356 sessions were collected in total, only 119 sessions were valid, with all sensors recorded properly and synchronized correctly. Participant position was also labelled every second over the duration of the session as one of: sitting up, left/right lateral, prone, supine. We divide the 119 valid sessions into train, validation and test sets according to a 60:10:30 ratio, based on participant id, with each session split into overlapping 60s segments (see Section~\ref{sec:sig_proc}).
        
        The ground truth HR was derived based on the R-R intervals detected in the ECG, and samples with high R-peak noise were rejected. We rejected 15\%, 10\% and 17\% of all segments in train, validation and test sets, respectively, due to the lack of reliable HR labels. The final dataset contained 168,496 overlapping 60s segments.
        \subsubsection{IR-UWB radar dataset}
        The data collection was performed at 3 sites: Savannah, Georgia, USA; Mobile, Alabama, USA; and Sheffield, South Yorkshire, England using an NXP SR160 UWB radar device (NXP Semiconductors, Eindhoven, Holland) fixed on an evaluation board. This UWB radar operates at a frequency of 8 GHz with one transmitting and two receiving antennas, and transmits pulse signals with a bandwidth of 0.5 GHz. The raw UWB signal was collected at a sample rate of 200 Hz.
        
\begin{figure}[h]
            \centering
        
            \sbox{\mybox}{\fbox{\includegraphics[height={4cm}]{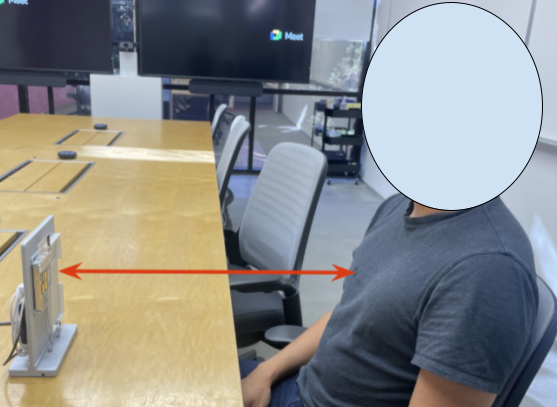}}}
            \begin{subfigure}[t]{\wd\mybox}
                \usebox{\mybox}
                \subcaption{Radar on the table \newline in front of the participant.}
                \label{fig:fig1_a}
            \end{subfigure}
            %
            \sbox{\mybox}{\fbox{\includegraphics[height={4cm}]{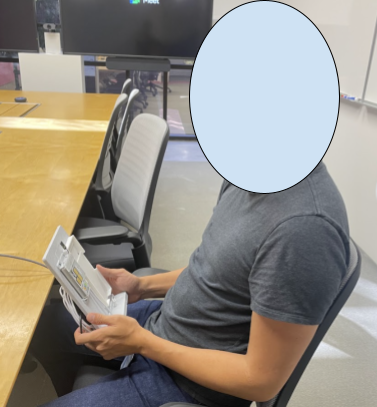}}}
            \begin{subfigure}[t]{\wd\mybox}
                \usebox{\mybox}
                \subcaption{Radar on the \newline participant's lap.}
                \label{fig:fig1_b}
            \end{subfigure}
            %
            \sbox{\mybox}{\fbox{\includegraphics[height={4cm}]{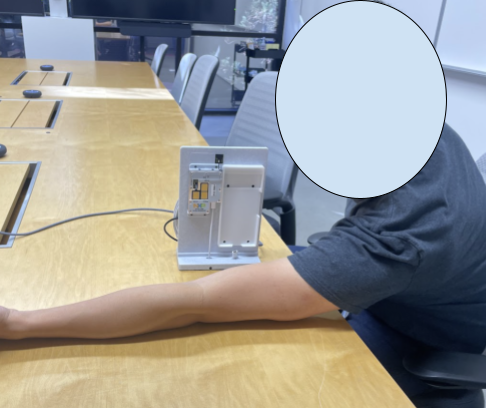}}}
            \begin{subfigure}[t]{\wd\mybox}
                \usebox{\mybox}
                \subcaption{Radar on the table \newline pointing at participant's upper arm.}
                \label{fig:fig1_c}
            \end{subfigure}
        
            \caption{Three radar positions used during the IR-UWB data collection study.}
            \label{fig:fig1}
        \end{figure}

        376 participants were recruited, of whom approximately 50\% had a confirmed diagnosis of diabetes or hypertension. Ground truth HR was derived from either ECG lead I or PPG signals of a BiosignalsPlux system (PLUX Biosignals, Lisbon, Portugal), a research-grade physiological data acquisition device. Two-minute recordings were captured from the UWB radar evaluation board in 3 positions: 1) on the table, approximately 2-3 feet in front of participant (to collect signals from head, neck and chest), 2) in the participant's lap (to collect signals from the same region but at a smaller distance), and 3) on the table pointed at the participant's upper arm (to collect signals from the brachial artery) (see Figure~\ref{fig:fig1}). In our study, we used data from all positions for training, but excluded the third position from validation and test sets as the signal was out-of-distribution compared to the mm-wave FMCW. This dataset was split into train, validation and test set according to a 50:20:30 ratio, again split on participant id, with each session split into overlapping 60s segments (see Section~\ref{sec:sig_proc}).
        
        The ground truth HR was calculated based on the ECG signal if it was reliable, or PPG otherwise, based on the R-R interval as described above. If both ECG and PPG signals were invalid due to high R-peak noise, the segment was rejected. We excluded 17\%, 13\% and 13\% of all segments in train, validation and test sets, respectively, due to the lack of reliable HR labels. The final dataset contained 4,696 overlapping 60s segments.
        
    \subsection{Signal Processing}
    \label{sec:sig_proc}
        \begin{figure}[h]
          \centering
          \fbox{\includegraphics[width=0.95\textwidth]{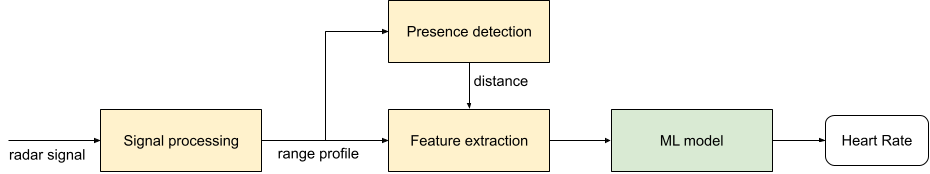} }
          
          \caption{The high-level architecture of the system, showing the signal processing, presence detection and feature extraction stages, followed by the ML model.}
          \label{fig:fig2}
        \end{figure}
        \subsubsection{mm-wave FMCW}
        We first averaged the chirps within each burst, resulting in a raw FMCW radar signal with a 30 Hz sampling rate and 256 samples per chirp. Then we split the data into overlapping 60s segments, with a step size of 15s, resulting in a data cube with dimensions 256 x 3 x 1800, where 256 is the number of “fast-time” samples per chirp, 3 is a number of receiving antennas and 1800 is the number of “slow-time” bursts within a 60s segment. For each 60s segment, we first applied a clutter filter by subtracting the average “slow-time” signal from the original signal. We then computed a fast Fourier transform (FFT) along the "fast-time" axis, to calculate a range profile with dimensions of 129 x 3 x 1800, where 129 is the number of range bins, 3 is the number of receiving antennas and 1800 is the "slow-time" axis. Note that this range profile is complex-valued.
        \subsubsection{IR-UWB}
        The raw IR-UWB signal is already represented as a range profile, which we downsampled via averaging from 200 Hz to 30 Hz to match the FMCW radar signal. Similarly to the previous dataset, the data was then split into overlapping 60s segments, with a step size of 15s, resulting in a data cube with dimension 52 x 2 x 1800, where 52 is the number of range bins, 2 is a number of receiving antennas and 1800 is the number of timesteps within that 60s segment.

    \subsection{Presence Detection}
    The first stage of our system is presence detection, to reduce and fix the number of range bins ingested to the model. Given a complex-valued range profile we performed a clutter filter as described above, followed by application of the constant false alarm rate (CFAR) algorithm \cite{ref_83} with a threshold of 1.5 to detect the range bin index with the highest power. We consider the resulting range bin index to be the participant’s location; if no such range bin is found, i.e. the participant cannot be detected, the segment was rejected.   
    
    \subsection{Feature Extraction}
        \subsubsection{mm-wave FMCW}
        We first extracted the complex-valued range profile for the range bin in which the participant was detected, and a number of extra bins around it. For mm-wave FMCW radar, 32 range bins around the user were extracted, covering a range of 0.86m. In line with prior work \cite{ref_1, ref_2, ref_3, ref_7, ref_62}, we considered the unwrapped representation of the range profile phases such that $2\pi$ signal discontinuities are removed. Building on this prior work, we also showed empirically that overall performance was boosted if we additionally considered the magnitude of the complex range profile, so for each range bin selected, the magnitude and unwrapped angle were extracted as two separate signals, each then filtered using a high pass filter to eliminate low-frequency noise below 3 Hz.
        
        To separate breathing motion and other body and environmental noise from the heart rate signal, we used an adaptive filter \cite{ref_50}. We first derived the respiratory trend from the signal by using a 3rd order Savitzky-Golay filter with a filter window length of 1.5 seconds, and then subtracted this trend from the original signal; this removed both the dominant and first harmonic frequency of the respiratory rate from the signal and rendered the heart rate much easier to extract. This differs from prior work that relies solely on band-pass filters to focus on the range of frequencies relevant to heart rate \cite{ref_61, ref_63, ref_25}, as these methods only filter out the fundamental respiratory rate frequency but retain its harmonics.
        The adaptive filter was applied on both the magnitude and unwrapped angle of each extracted range bin from the range profile, from each receiving antenna.

        \subsubsection{IR-UWB}
        For IR-UWB radar, a single range bin where the user was detected was extracted, which covers a range of 0.3m. For the IR-UWB radar signal we also modeled both the magnitude and phase components of the complex range-profile from the extracted range bin as two separate signals, in contrast to prior work that focuses on only considering the magnitude \cite{ref_4, ref_5, ref_6, ref_62}. The same high-pass and adaptive filters as described above were then applied to the signal.

    \subsection{Modeling}
    
        \begin{figure}[bt]
          \centering
          \fbox{\includegraphics[width=0.95\textwidth]{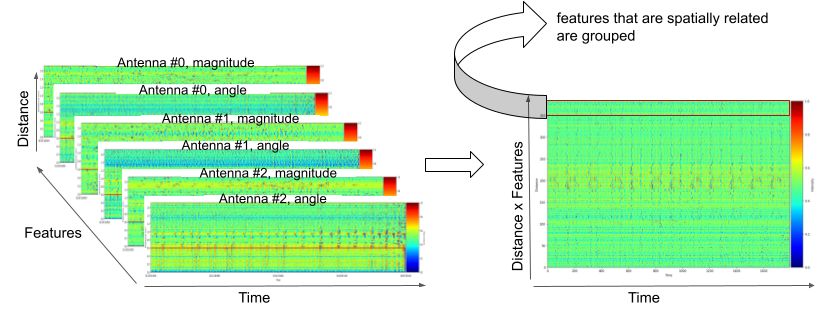} }
          \caption{Flattening of the mm-wave FMCW data that is performed in the Input Layer. Features that are spatially related are grouped together.}
          \label{fig:fig3}
        \end{figure}

    As described below, we explored the best possible model trained from scratch on mm-wave FMCW and IR-UWB radar datasets separately, as well as the most transferable model from mm-wave FMCW to IR-UWB radar (see Section~\ref{sec:transfer_learning}).
        \subsubsection{Model Architecture}
\begin{figure}[h]
          \centering
          \fbox{\includegraphics[width=0.95\textwidth]{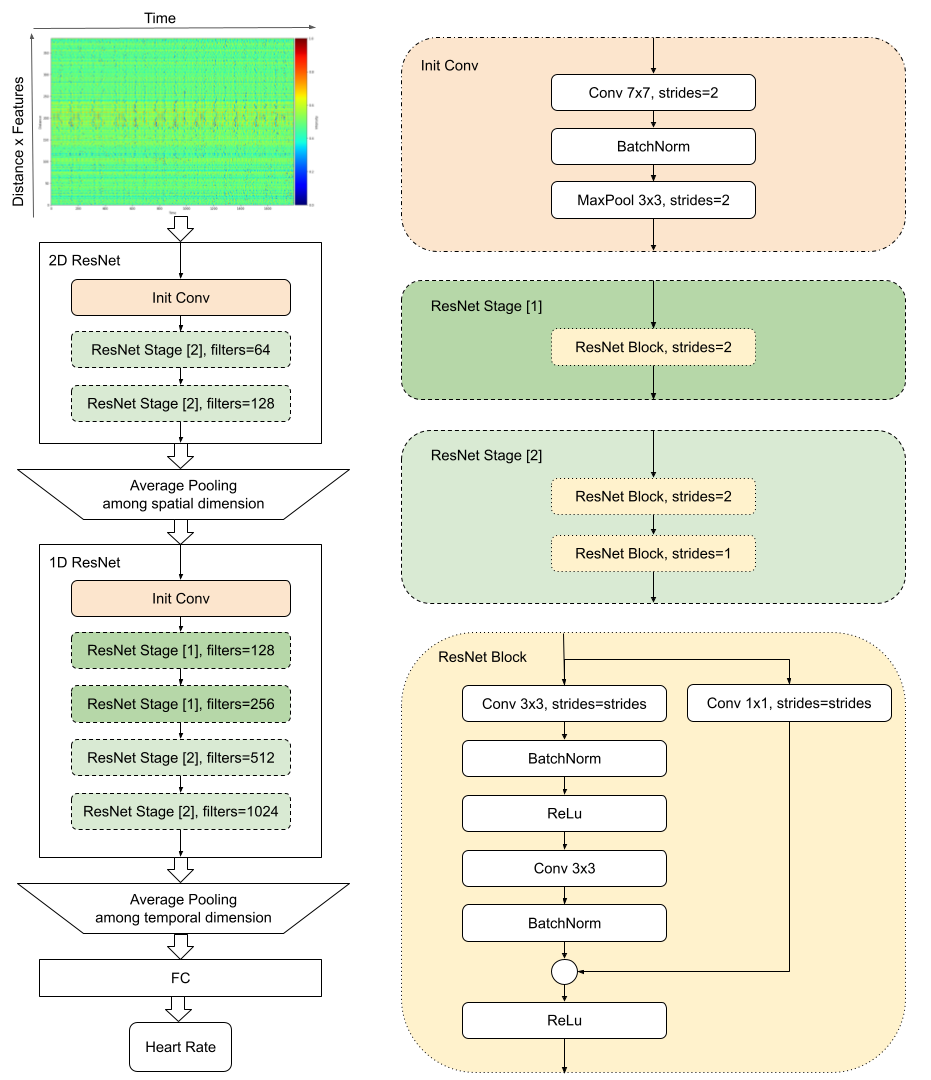} }
          \caption{Novel model architecture which combines a 2D and 1D ResNet network.}
          \label{fig:fig4}
        \end{figure}

        We used a 5-layer novel model architecture which combines a 2D and 1D ResNet network (see Figure~\ref{fig:fig4}).
        \begin{enumerate}
            \item \textbf{Input layer: }
            The preprocessed features (described earlier) were reshaped to $T$ x $S$ x $1$, where $T$ is the time dimension, $S = 2 \cdot num_{antennas} \cdot num_{range\_bins}$ is the flattened spatial dimension which consists of both the magnitude and unwrapped angle per each $num_{antennas}$ antenna per each $num_{range\_bins}$ range bins, and $1$ is the number of channels. This ensures that the features that are spatially related are grouped together (see Figure~\ref{fig:fig3}). Each feature was scaled using min-max normalization.
            \item \textbf{2D ResNet: }
            We first used a 2D ResNet because both spatial and temporal relationships between features are useful in the early layers, as spatially close features reflect the temporal micromotions from neighbouring parts of the body. The 2D ResNet consists of 2 stages with $64$ and $128$ filters respectively, and a stride of 2 to increase the size of the receptive fields and enable feature hierarchy. Each stage consists of up to two ResNet blocks. The output of the 2D ResNet was a 3D data cube of size $T/8$ x $S/8$ x $128$, where temporal and spatial dimensions respectively are smaller than the original input size due to striding and pooling, and $128$ is the number of features. 
            \item \textbf{Average pooling: }
            By this stage in the network, the features that emerged from spatially correlated signals were already extracted, so we collapsed the spatial dimension via averaging, resulting in a 2D data cube of size $T/8$ x $128$.
            \item \textbf{1D ResNet: }
            The result from the previous stage was fed into a 1D ResNet, which extracted more specific features in the increasingly larger receptive field along the temporal dimension. It consists of 4 stages with $128$, $256$, $512$ and $1024$ filters respectively, and a stride of 2. The resulting data cube had size of $T/(8 \cdot 32)$ x $1024$, where a temporal dimension is smaller than the input size due to striding and pooling, and $1024$ is the number of features. At this point, the features for all timesteps were averaged, resulting in a vector of size $1024$.
            \item \textbf{Output layer: }
            One fully connected layer was then used to estimate the HR from extracted features. 
        \end{enumerate}
        \subsubsection{Model Training FMCW radar}
        The original model was trained on mm-wave FMCW radar data using the AdamW optimizer \cite{ref_81} for 200k steps with the batch size of 32, and a constant learning rate of 0.001. We performed hyperparameter sweeps focused on the number and sizes of stages for both 2D and 1D ResNets.

        We report accuracy on the mm-wave FMCW dataset using this model.
        \subsubsection{Transfer Learning}
        \label{sec:transfer_learning}
        
        Mm-wave FMCW and IR-UWB radars have very different antenna and bandwidth characteristics, as shown in Table~\ref{tab:tab1}, resulting in UWB having a ~10x lower range resolution of 0.3 m compared to FMCW’s resolution of 0.027 m.

        To overcome the challenges this signal difference poses to model transfer, we performed additional preprocessing steps to modify the mm-wave FMCW radar data to better resemble the target IR-UWB data, effectively lowering its range resolution. In particular, we selected a single range bin from a single receiving antenna. We then applied data augmentation to add Gaussian noise with zero mean and standard deviation of 0.0005, with probability of 0.7. We re-trained our model from scratch on this alternately preprocessed FMCW data.
        
        \begin{table}[t]
         \caption{Specification of mm-wave FMCW and IR-UWB radars used in this study.}
          \centering
          \begin{tabular}{l p{5.5cm} p{5.5cm}}
            \toprule
            \textbf{} & \textbf{Mm-wave FMCW Radar\newline(Soli radar in Nest Hub)} & \textbf{IR-UWB Radar\newline(Evaluation board)} \\
            \midrule
            \textbf{Transmitted signal} & Chirp - continuous sinusoidal wave whose frequency increases linearly with time & Very short pulses with duration in the order of a few nanoseconds to a few hundred picoseconds \\
            \textbf{Signal type} & Instantaneous Frequency (real) & Range profile (complex) \\
            \textbf{Frequency} & 60 GHz & 8 GHz \\
            \textbf{Bandwidth} & 5.5 GHz & 500 MHz \\
            \textbf{Range resolution} & 2.7 cm & 30 cm \\
            \textbf{Receiving antennas} & 3 & 2 \\
            \bottomrule
            \end{tabular}
          \label{tab:tab1}
        \end{table}
        
        \begin{figure}[t]
          \centering
          \fbox{\includegraphics[width=0.95\textwidth]{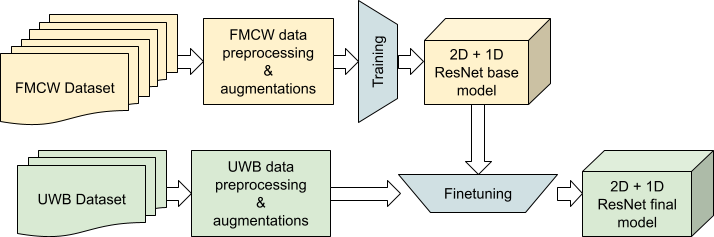} }
          \caption{High level architecture of model transfer.}
          \label{fig:fig5}
        \end{figure}
        \subsubsection{Model Training and Fine-tuning IR-UWB radar}
        \label{sec:UWB_taining_finetuning}
        The model was re-trained using AdamW optimizer \cite{ref_81} for 140k steps with the batch size of 64, and a constant learning rate of 0.001. The model was selected based on the validation set performance, and the best model achieved an MAE of 2.450 bpm.
        
        We then applied similar preprocessing steps to IR-UWB data, selecting a single range bin from a single antenna. Additionally, we swapped the order of the features (magnitude and unwrapped angle).
        
        Finally, we fine-tuned the model pre-trained using FMCW data on the preprocessed IR-UWB data. During the fine-tuning process, we also applied data augmentations, by adding Gaussian noise to the data with the same characteristics and probability of 0.6. The model was trained using AdamW optimizer \cite{ref_81} for 2.7k steps with the batch size of 2048, and an exponential decay learning schedule with the learning rate of 0.0003 and decay rate of 0.1.
        
        To establish a baseline, the same model architecture was also trained on the IR-UWB dataset from scratch, without leveraging any transfer learning techniques. The best baseline model was trained on antennas split into separate examples, and a single range bin. The model was trained using AdamW optimizer \cite{ref_81} for 55k steps with the batch size of 2048, and a constant learning rate of 0.001.
        \subsubsection{Ablation Experiments}     
        In the Supplementary Information section B2, we describe ablation experiments that verify our proposed model architecture, feature selection, data augmentation, and preprocessing choices.
        \subsubsection{Statistical Analysis}
        For our results, we report Mean Absolute Error (MAE) and Mean Absolute Percentage Error (MAPE). To produce the 95\% confidence interval we make use of bootstrapping with 1,000 replicates. We provide Bland-Altman analysis to assess the agreement between predicted HR the ground truth HR values \cite{ref_94}. All metrics are reported on the test set.

%
%
\section{Results}
    Below we report the performance of our novel model on mm-wave FMCW radar and its performance on IR-UWB radar after transfer learning.
    \subsection{mm-wave FMCW radar}
    As described in the Section~\ref{sec:methods}, the mm-wave FMCW radar dataset originates from an overnight sleep study \cite{ref_7} with a Google Nest Hub device placed by the bedside (total N=199 valid sessions, 168,496 overlapping 60-second segments).

    The presence detection block detected the user in 98.9\% of test segments (test recall 98.9\%). The model performance below is reported on those segments where the user was detected.
    
    With an MAE of 0.85 bpm and MAPE of 1.42\%, our approach substantially improved on the previous SOTA on this dataset (MAE 1.69 bpm, MAPE 2.67\%, 88.53\% recall)\cite{ref_7}. As a reference, ANSI/AAMI standards for heart rate measurement for consumer devices require an accuracy of up to 5 bpm MAE and 10\% MAPE \cite{ref_58, ref_59, ref_60}.
    
    As shown in Table~\ref{tab:tab2}, the model performs well across most body positions and maintains good performance even when the person is moving between positions. The model also performs well across HR bands up to 100 bpm, reflecting the heart range rate in this sleep dataset, and for distances up to 2m, corresponding to the typical distance between the sleeper and the bedside table (Table~\ref{tab:tab3} and Table~\ref{tab:tab4}).
    
    Through ablation experiments (~\ref{sec:FMCW_ablations}), we demonstrate that our chosen model architecture (2D~+~1D ResNet, ~\ref{tab:tabs1}), including both unwrapped angle and magnitude as features (~\ref{tab:tabs2}), and concatenating multiple antennas (~\ref{tab:tabs3}) all contribute to the high accuracy of this model.
    
    \begin{table}[H]
     \caption{Model performance on mm-wave FMCW radar data including 95\% confidence interval by body position. The "Moving" label is used if the participant changed their position during the 60s segment.}
      \centering
      \begin{tabular}{lcc}
        \toprule
        \textbf{Body Pose}      & \textbf{Test MAE [95\% CI]} & \textbf{\% of test samples} \\
        \midrule
        Overall         & 0.85 [0.83, 0.87]         & $-$                           \\
        \midrule
        Prone           & 0.43 [0.39, 0.47]         & 3.4                         \\
        Supine          & 1.04 [1.00, 1.07]         & 39.3                        \\
        Left lateral   & 0.80 [0.74, 0.86]         & 19.8                        \\
        Right lateral  & 0.60 [0.58, 0.62]         & 31.2                        \\
        Sitting up     & 4.86 [3.27, 6.40]         & 0.2                         \\
        Moving          & 1.24 [1.16, 1.33]         & 6.0                         \\
        \bottomrule
        \end{tabular}
      \label{tab:tab2}
    \end{table}
    
    \begin{table}[H]
     \caption{Model performance on mm-wave FMCW radar data including 95\% confidence interval by HR bands.}
      \centering
      \begin{tabular}{lcc}
        \toprule
        \textbf{HR band} & \textbf{Test MAE [95\% CI]} & \textbf{\% of test samples} \\
        \midrule
        Overall          & 0.85 [0.83, 0.87]         & $-$                         \\
        \midrule
        \text{[0, 50)}         & 2.42 [2.10, 2.75]         & 3.4                         \\
        \text{[50, 60)}        & 0.72 [0.69, 0.75]         & 40.1                        \\
        \text{[60, 70)}        & 0.88 [0.85, 0.91]         & 38.3                        \\
        \text{[70, 80)}       & 0.77 [0.74, 0.81]         & 15.4                        \\
        \text{[80, 90)}        & 0.86 [0.77, 0.96]         & 2.7                         \\
        \text{[90, 100)}       & 0.81 [0.42, 1.40]         & 0.1                         \\
        \bottomrule
        \end{tabular}
      \label{tab:tab3}
    \end{table}

    \begin{table}[H]
     \caption{Model performance on mm-wave FMCW radar data including 95\% confidence interval by user distance from the radar.}
      \centering
      \begin{tabular}{lcc}
        \toprule
        \textbf{User distance} & \textbf{Test MAE [95\% CI]} & \textbf{\% of test samples} \\
        \midrule
        Overall                & 0.85 [0.83, 0.87]         & $-$                         \\
        \midrule
        \text{[0m, 0.5m)}   & 4.74 [4.28, 5.24]         & 2.4                         \\
        \text{[0.5m, 1.0m)} & 0.76 [0.74, 0.78]         & 57.4                        \\
        \text{[1.0m, 1.5m)} & 0.72 [0.70, 0.75]         & 40                          \\
        \text{[1.5m, 2.0m)} & 5.04 [3.69, 6.33]         & 0.2                         \\
        \text{[2.0m, 2.5m)} & 6.00 [1.41, 10.38]        & 0.01                        \\
        \bottomrule
        \end{tabular}
      \label{tab:tab4}
    \end{table}
    
    \begin{figure}[ht]
      \centering
      \fbox{\includegraphics[width=0.8\textwidth]{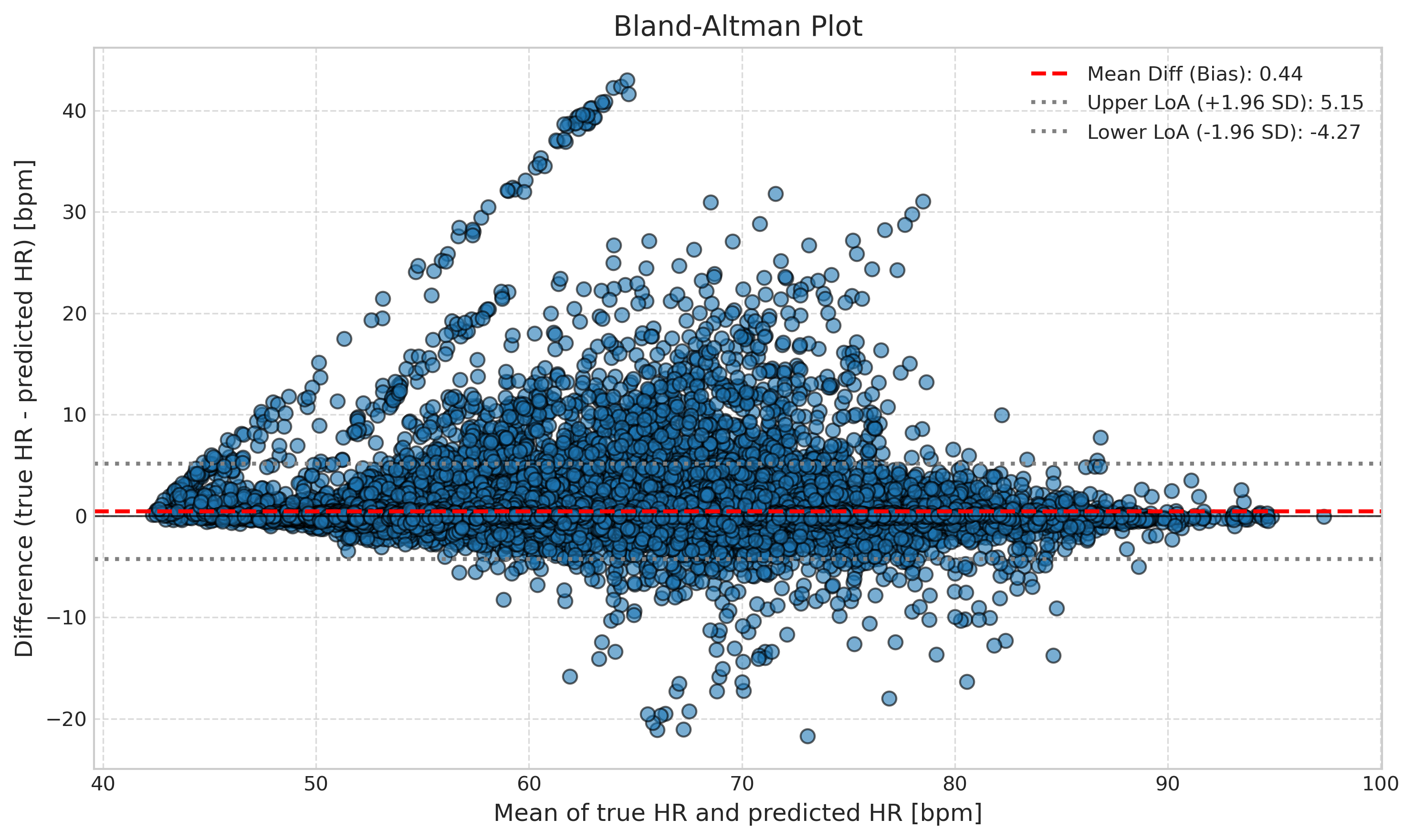} }
      \caption{Bland-Altman analysis to assess the agreement between predicted HR and the ground truth HR values on FMCW radar data.}
      \label{fig:fig6}
    \end{figure}
    
    \begin{figure}
      \centering
      \fbox{\includegraphics[width=0.95\textwidth]{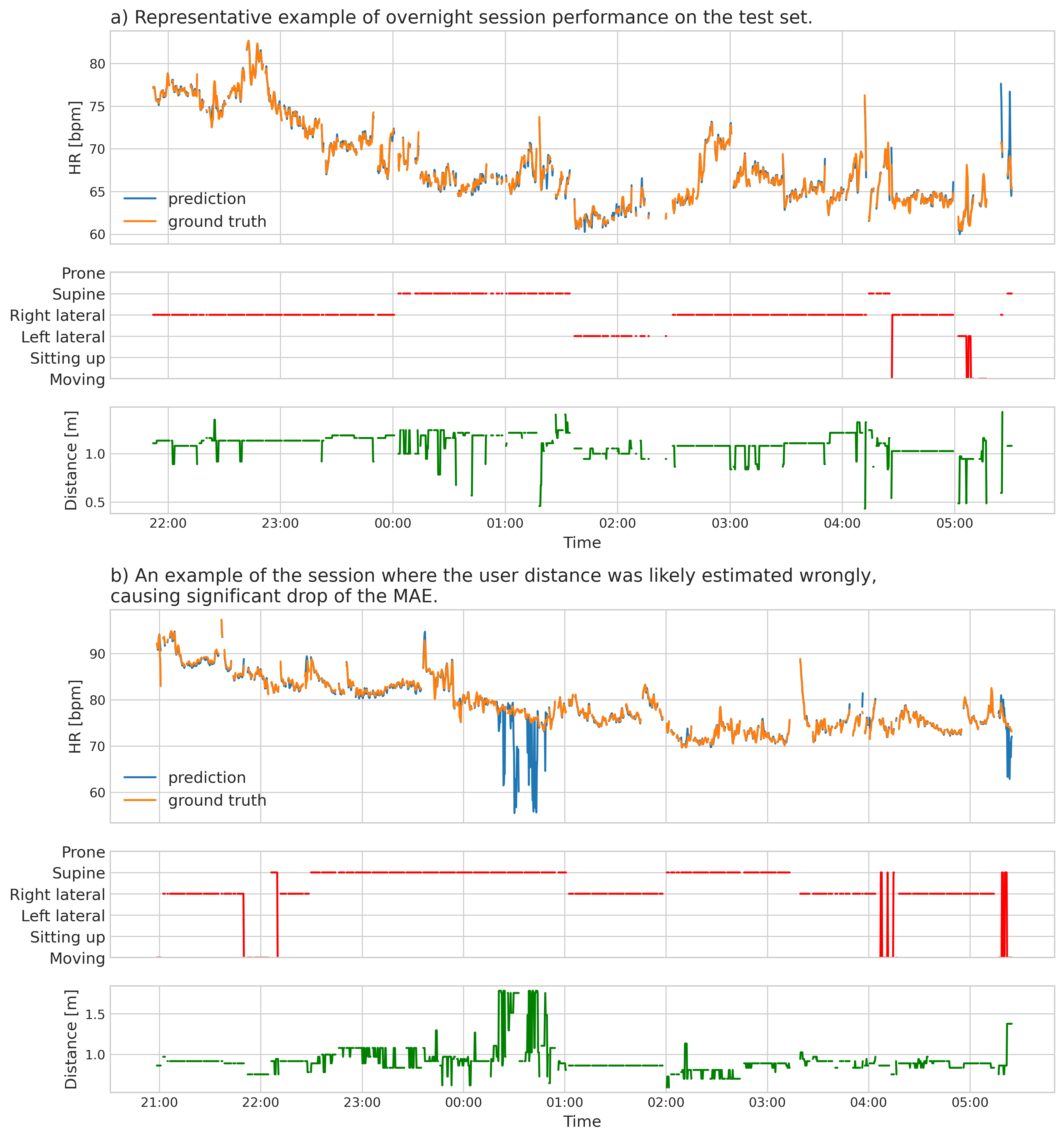} }
      \caption{a) Representative example of overnight session performance on the test set. b) An example of the session where the user distance was likely estimated wrongly as being above 1.5 meter between midnight and 1:00 am (where the true distance was likely to be around 1.0 meter from the radar), causing significant drop of the MAE for those segments. Within each subfigure, the top plot shows the model performance (blue) compared to the ground truth (orange), the middle plot shows the body position and the bottom plot shows the estimated user distance from the radar.}
      \label{fig:fig7}
    \end{figure}
    \subsection{Transfer learning and IR-UWB radar}
    The IR-UWB radar dataset is much smaller (376 sessions, 4,696 overlapping 60-second segments) than the FMCW dataset described above. The radar configuration (Table~\ref{tab:tab1}), the position and the distance of the device relative to the subject differed from the FMCW dataset (Figure~\ref{fig:fig1}). Despite these differences we show that accuracy of heart rate measurement on this dataset improves after transfer learning of the features learned from the FMCW radar.

    The presence detection block detected the user in 97.5\% of test segments (test recall 97.5\%). The model performance below is reported on those segments where the user was detected.
    
    As a baseline, we trained our model from scratch using IR-UWB radar data, using the same architecture as that used for the FMCW dataset (~\ref{sec:UWB_taining_finetuning}). The best baseline model achieved an MAE of 5.4 bpm and MAPE of 8.4\% on the test set.
    
    To achieve the most transferable model using the FMCW dataset, we re-trained the model after additional preprocessing steps to modify the mm-wave FMCW radar data to better resemble the target IR-UWB data, effectively lowering its range resolution (Section~\ref{sec:transfer_learning}). This model achieves an MAE of 2.45 bpm on the FMCW test dataset. 
    
    We then fine-tuned this model on the IR-UWB dataset, achieving an MAE of 4.1 and MAPE of 6.3\% (Table~\ref{tab:tab5}, Figure~\ref{fig:fig9}). This amounts to a 25\% improvement compared to the best model trained from scratch on IR-UWB data alone and meets the recommended 5 bpm MAE and 10\% MAPE heart rate accuracy threshold for consumer devices \cite{ref_58, ref_59, ref_60}. Our experiments show that using the first 40\% of the training set brings the most significant model improvements, beyond which point the gain is much smaller (~\ref{sec:UWB_train_set_size}). 
    
    We show that model performance holds for both positions of the device relative to the user (Table~\ref{tab:tab6}). These positions correspond to common locations that individuals might hold their mobile phone when in use. Model accuracy is higher for heart rates that are well-represented in the training dataset (Table~\ref{tab:tab7}). There is no statistically significant difference in performance based on participant's underlying cardiovascular disease or diabetes (Table~\ref{tab:tab9}), nor by their age or gender (Tables~\ref{tab:tab10} and Table~\ref{tab:tab11}). Performance varies by data collection site, indicating that the model may be sensitive to small changes in device configuration (Table~\ref{tab:tab8}).  
    
    Our ablation experiments show that, as with FMCW, including both unwrapped angle and magnitude as features leads to a performance gain for IR-UWB (~\ref{tab:tabs5}). Training and fine-tuning the model on a single antenna and a single range bin leads to the best performance, and adding data augmentation further boosts the performance. Additionally, we also show that swapping the order of the features (magnitude and unwrapped angle) boosts performance, as the relative significance of these features depends on the radar type (see ablations in ~\ref{sec:UWB_augmentations}). Finally, we show that while a standard 1D ResNet model architecture can reach a similar performance, our proposed 2D~+~1D ResNet achieves the same performance with 15x fewer parameters (~\ref{sec:UWB_model_architecture}).
    
    \begin{figure}[h]
      \centering
      \fbox{\includegraphics[width=0.8\textwidth]{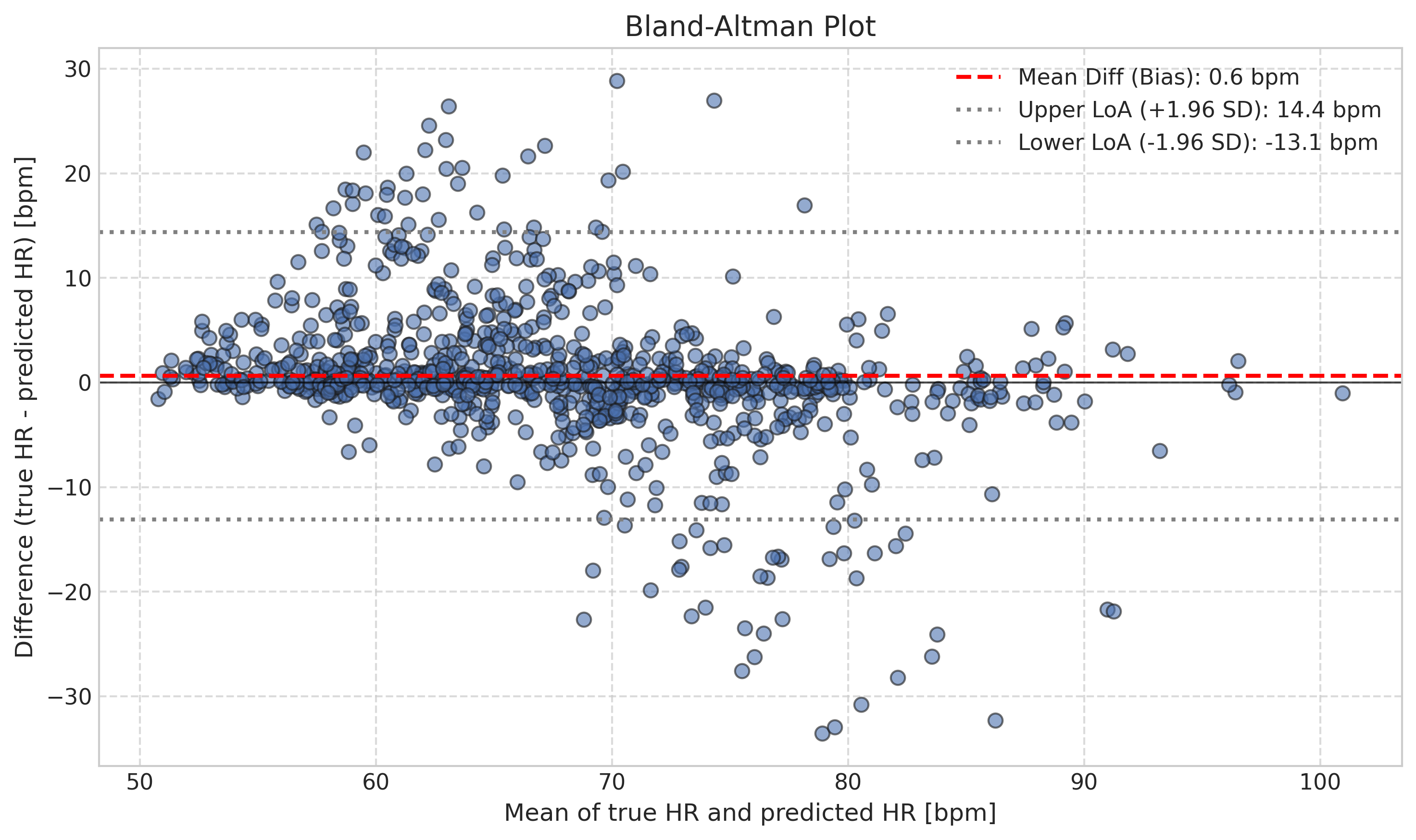} }
      \caption{Bland-Altman analysis to assess the agreement between predicted HR the ground truth HR values on IR-UWB radar data.}
      \label{fig:fig8}
    \end{figure}
    
    \begin{table}[H]
     \caption{Comparison of the baseline and transfer learning models performance on IR-UWB radar data including MAE with 95\% confidence interval and MAPE.}
      \centering
      \begin{tabular}{lcc}
        \toprule
        \textbf{Model}          & \textbf{Test MAE [95\% CI]} & \textbf{Test MAPE} \\
        \midrule
        Baseline model    & 5.4 [5.0, 5.8]            & 8.4                \\
        Transfer learning & 4.1 [3.8, 4.5]            & 6.3                \\
        \bottomrule
        \end{tabular}
      \label{tab:tab5}
    \end{table}

    \begin{table}[H]
     \caption{Model performance on IR-UWB radar data including 95\% confidence interval by radar position.}
      \centering
      \begin{tabular}{lcc}
        \toprule
        \textbf{Position}                        & \textbf{Test MAE}   & \textbf{\% of test samples} \\
        \midrule
        Overall                                  & 4.1 [3.8, 4.5]      & $-$                         \\
        \midrule
        On the table in front of participant & 4.3 [3.8, 4.8]      & 49.1                        \\
        On participant's lap                   & 3.9 [3.5, 4.5]      & 50.9                        \\
        \bottomrule
      \end{tabular}
      \label{tab:tab6}
    \end{table}
    
    \begin{table}[H]
     \caption{Model performance on IR-UWB radar data including 95\% confidence interval by HR bands.}
      \centering
      \begin{tabular}{lcc}
        \toprule
        \textbf{HR band}  & \textbf{Test MAE [95\% CI]} & \textbf{\% of test samples} \\
        \midrule
        Overall           & 4.1 [3.8, 4.5]            & $-$                         \\
        \midrule
        \text{[0, 60)}          & 5.1 [4.4, 5.9]            & 28.4                        \\
        \text{[60, 70)}         & 3.0 [2.6, 3.5]            & 32.2                        \\
        \text{[70, 80)}         & 2.5 [2.1, 2.8]            & 28.2                        \\
        \text{[80, 90)}         & 7.7 [6.1, 9.5]            & 9.2                         \\
        \text{[90, 100)}        & 13.9 [7.6, 20.4]          & 1.6                         \\
        \text{[100, 110)}       & 19.2 [6.2, 29.7]          & 0.4                         \\
        \bottomrule
        \end{tabular}
      \label{tab:tab7}
    \end{table}
    
    \begin{table}[H]
     \caption{Model performance on IR-UWB radar data including 95\% confidence interval by site.}
      \centering
      \begin{tabular}{lcc}
        \toprule
        \textbf{Site}          & \textbf{Test MAE [95\% CI]} & \textbf{\% of test samples} \\
        \midrule
        Overall          & 4.1 [3.8, 4.5]            & $-$                         \\
        \midrule
        Mobile (USA)     & 5.7 [4.4, 7.3]            & 11.5                        \\
        Savannah (USA)   & 4.4 [4.0, 4.8]            & 53.5                        \\
        Sheffield (UK)   & 3.0 [2.4, 3.7]            & 31                          \\
        \bottomrule
        \end{tabular}
      \label{tab:tab8}
    \end{table}
    
    \begin{table}[H]
     \caption{Model performance on IR-UWB radar data including 95\% confidence interval by condition.}
      \centering
      \begin{tabular}{lcc}
        \toprule
        \textbf{Condition}                        & \textbf{Test MAE [95\% CI]} & \textbf{\% of test samples} \\
        \midrule
        Overall                                 & 4.1 [3.8, 4.5]            & $-$                         \\
        \midrule
        Cardiovascular Disease                  & 4.0 [3.5, 4.6]            & 40.2                        \\
        Cardiovascular Disease and Diabetes     & 5.2 [4.1, 6.4]            & 16.1                        \\
        Diabetes                                & 1.4 [0.5, 2.4]            & 1.0                         \\
        None                                    & 4.0 [3.4, 4.5]            & 38.8                        \\
        Unknown                                 & 3.0 [2.1, 4.1]            & 3.9                         \\
        \bottomrule
      \end{tabular}
      \label{tab:tab9}
    \end{table}
    
    \begin{table}[H]
     \caption{Model performance on IR-UWB radar data including 95\% confidence interval by age.}
      \centering
      \begin{tabular}{lcc}
        \toprule
        \textbf{Age}      & \textbf{Test MAE [95\% CI]} & \textbf{\% of test samples} \\
        \midrule
        Overall     & 4.1 [3.8, 4.5]            & $-$                         \\
        \midrule
        \text{[40, 50)}    & 3.8 [2.8, 4.6]            & 16.8                        \\
        \text{[50, 60)}   & 4.0 [3.4, 4.6]            & 29.8                        \\
        \text{[60, 70)}   & 4.5 [3.9, 5.1]            & 37.9                        \\
        \text{[70, 80+)}  & 3.9 [3.2, 4.7]            & 15.5                        \\
        \bottomrule
        \end{tabular}
      \label{tab:tab10}
    \end{table}
    
    \begin{table}[H]
     \caption{Model performance on IR-UWB radar data including 95\% confidence interval by gender.}
      \centering
      \begin{tabular}{lcc}
        \toprule
        \textbf{Gender} & \textbf{Test MAE [95\% CI]} & \textbf{\% of test samples} \\
        \midrule
        Overall   & 4.1 [3.8, 4.5]            & $-$                         \\
        \midrule
        Female    & 4.4 [3.9, 4.9]            & 74                          \\
        Male      & 3.5 [2.9, 4.1]            & 26                          \\
        \bottomrule
        \end{tabular}
      \label{tab:tab11}
    \end{table}
    
    \begin{figure}[htp]
      \centering
      \fbox{\includegraphics[width=0.95\textwidth]{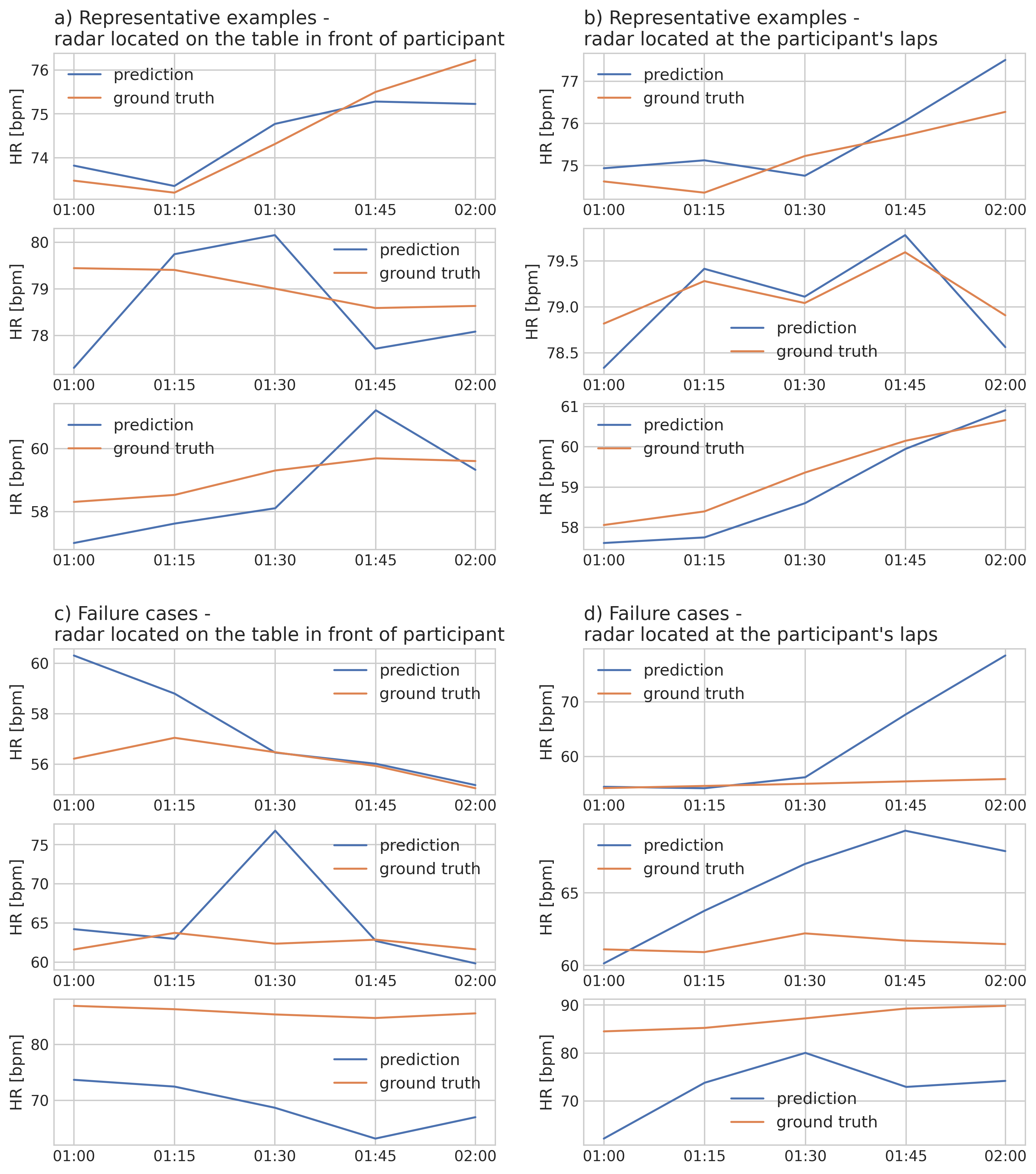} }
      \caption{a), b) Representative examples of the model performance (blue) compared to the ground truth (orange) for 3 selected participants from the test set for the session where the radar was located on the table in front of the participant (a) and at the participant's lap (b). c), d) Failure cases of the model performance (blue) compared to the ground truth (orange) for 3 selected participants from the test set for the session where the radar was located on the table in front of the participant (c) and at the participant's lap (d).}
      \label{fig:fig9}
    \end{figure}

%
%
\section{Discussion}
    \subsection{Discussion of methods and experimental results}
    The radar signal contains unique information in the spatial and temporal dimensions, and one of the primary contributions of our work is to design a model architecture that leverages these. Through ablation experiments ( ~\ref{sec:FMCW_model_architecture}), we demonstrate that our proposed 2D~+~1D model architecture outperforms other standard model architectures by a large margin with a 0.89 bpm validation set MAE versus an 1.30 bpm MAE of when using a standard 1D ResNet per feature.
    
    Using the same 2D~+~1D ResNet model architecture also brings the benefits for the IR-UWB radar dataset (~\ref{sec:UWB_model_architecture}), despite the fact that only two features are used as an input to the model: unwrapped angle and magnitude from a single antenna and a single range bin. Achieving a similar performance using the standard 1D ResNet required over 15x more parameters. These experiments demonstrate the value of capturing both spatial and temporal relationships between features in early model stages.
    
    The user detection and range bins extraction block represents the primary area for improvement in our current system. We use a relatively simple constant false alarm rate (CFAR) algorithm to detect the peak in the power range profile and extract a fixed number of range bins. The extracted range bins sometimes fail to include the participant, causing a sharp increase in MAE, as mentioned below. Alternative user detection methods that dynamically select the range bins around the user, e.g. by leveraging ML techniques, could improve overall performance.
    
    To better understand model performance and failure modes, we performed subgroup analyses for both radar types, within the constraints of the study design. On the FMCW dataset obtained by placing the sensor on the bedside during overnight sleep we observe that the model generalizes well to various sleep positions, and maintains good performance even when the person transitions between positions (see Table~\ref{tab:tab2}). The performance drop in the sitting-up position is attributable to the fact that a third of the sitting-up samples in the test set came from the very end of one of the sessions, where there was likely a lot of movement as the participant was leaving the bed.
    
    The performance of our model also drops slightly for the lowest distance bucket (0 - 0.5 m) and more substantially for distance > 1.5 m. However, after inspecting individual traces from test set sessions (see one example in Figure~\ref{fig:fig7} b), these drops are explained by a failure of the user detection algorithm at the preprocessing stage.
    
    Model performance is consistent across all heart range ranges on FMCW radar. However, as the dataset represents overnight sleep, there are only a handful of samples in the higher heart rate ranges (0.1\% with HR > 90 bpm). Higher heart rates are similarly underrepresented in the IR-UWB study (2\% with HR > 90 bpm) which collected resting HR while participants were seated, where we see a drop in performance for these heart rate bins. It is important to focus future work on better evaluating and optimizing performance at higher heart rates, which are expected during exercise, or in individuals with clinical illness. In addition to the obvious need for collection of datasets that include more samples with higher heart rates, our preliminary experiments show that applying data augmentation to accelerate HR can boost the performance for the high HR subgroups (see ~\ref{sec:UWB_high_hr} for details). In addition, our exploration of the impact of the presence of underlying cardiovascular disease and diabetes did not show convincing evidence of differential accuracy among people with these comorbidities, with the exception of lower MAE in the group of individuals with diabetes which was likely due to small sample size. Further research, with larger numbers of individuals with comorbidities such as these, could further explore whether accuracy of heart rate measurement differs in such individuals from those without these conditions. In addition, the potential for UWB to be able to detect vascular changes (e.g. arterial stiffness) known to occur with conditions such as hypertension and diabetes could be valuable clinically \cite{ref_88, ref_89, ref_90}. While beyond the scope of our work, we also believe that there is potential in using a similar model architecture to estimate Heart Rate Variability (HRV) or even reconstruct ECG or PPG pseudo-waveforms.
    
    The transfer learning techniques presented in this paper are appropriate for the specific types and configurations of the radar systems used in this study, and may vary for other types of radar, depending on their characteristics, including type of transmitted signal, bandwidth and number and quality of receiving antennas. More research is needed to develop and evaluate algorithms that work well in a more noisy environment, for example with multiple individuals in the room or during exercise. Our datasets did not represent these situations. While the FMCW dataset was collected using a Nest Hub device placed on the participants’ bedside, the IR-UWB radar study was performed under idealized laboratory conditions, using an NXP SR160 UWB sensor fixed onto an evaluation board, using a configuration similar to that available on mobile devices. Further work needs to be done to assess the performance of the model with the UWB radar sensor fully built into the consumer device, as the performance of the algorithm likely fluctuates based on the specific hardware and software configuration, and the quality of the received signal.
    \subsection{Clinical Implications}
    Measurement of heart rate is used routinely in most healthcare settings, either as an episodic ‘spot check’ (for example in outpatients) or for continuous monitoring (in hospitalised patients) \cite{ref_84, ref_85}. Such monitoring is proven to detect changes in health status including onset of common complications including heart rhythm disorders or infection \cite{ref_86, ref_87}. While there are several technologies available for measurement of heart rate in clinical settings, such as finger pulse oximetry or ECG, all involve direct contact with the individual. There are few contactless methods for heart rate measurement, other than simple observation used in routine clinical practice. While the current study focussed on adults, there is strong evidence for the value of heart rate monitoring in children for acute illness management \cite{ref_91, ref_92, ref_93}. Contactless methods of measurement of heart rate in children would provide even more value by avoiding the need for medical equipment to touch the child which can cause distress. However, not only would evidence of accuracy for faster heart rates be required, but also evidence to disambiguate the parent and child heart rates (given that sick children are often held by parents or health care staff in clinical settings).
    
    In addition to clinical settings, accurate measurement of heart rate could provide value for people at home. For some individuals this could be used as a spot check for heart rate when they are concerned about clinical illness of multiple types (in themselves or others), or during telemedicine consultations. The ability to potentially obtain contactless measurement of heart rate, as well as breathing rate, and temperature using smartphones could facilitate more accessible and ubiquitous vital sign measurement and could provide patients and healthcare systems with the ability to detect and intervene on deterioration in health much earlier than is currently possible. Wider use cases for contactless heart rate (and breathing rate) measurement among consumers could include monitoring response to exercise, meditation, as well as part of sleep monitoring - all without the need for wearable devices. As noted above, further research is needed however to provide evidence of accuracy at faster heart rates, impact of gross body movements (e.g while using exercise equipment) as well as numerous other environmental sources of disruption (e.g. other individuals, fans).

%
%
\section{Conclusion}
This work demonstrates that a novel model architecture, which combines 2D and 1D ResNet blocks, can be used to achieve low mean absolute error for heart rate detection from mm-wave FMCW radar by extracting both spatial and temporal features from range profile data. Furthermore, we demonstrate for the first time that transfer learning is feasible between mm-wave FMCW and IR-UWB radar types for HR monitoring, enabling clinical grade performance with a relatively small training dataset.

%
%
\section*{Acknowledgments}
We would like to thank Abhijit Guha Roy, Michał Matuszak and Florence Thng for helping review the paper and Shwetak Patel and Dale Webster for their support. We also thank Catherine Kozlowski for legal support. The mm-wave FMCW radar dataset collection was approved under IRB (no. Pro00042166). The IR-UWB dataset collection was approved under IRB (no. Pro00074869). The ethical approval to collect dataset at the Sheffield, South Yorkshire, England site was granted by the UK Health Research Authority (23/YH/0251).
    
    \bibliographystyle{unsrt}  
    \bibliography{references}  

    \begin{appendices}
    \renewcommand{\thesubsection}{\Alph{subsection}.}
    \renewcommand{\thesubsubsection}{\thesubsection\arabic{subsubsection}.}
    \setcounter{secnumdepth}{5}
    \renewcommand{\theparagraph}{\thesubsubsection\arabic{paragraph}.}
    \renewcommand{\thesubparagraph}{\theparagraph\arabic{subparagraph}.}
    \titleformat{\paragraph}{\normalfont\normalsize\bfseries}{\theparagraph}{1em}{}
    \titleformat{\subparagraph}{\normalfont\normalsize\bfseries}{\thesubparagraph}{1em}{}
    
    \setcounter{figure}{0}
    \setcounter{table}{0}
    \renewcommand{\thefigure}{S\arabic{figure}}
    \renewcommand{\thetable}{S\arabic{table}}
    \renewcommand{\figurename}{Supplementary Figure}
    \renewcommand{\tablename}{Supplementary Table}

%
%
\clearpage
\appendix
\section*{Supplementary Information}
    \subsection{mm-wave FMCW radar performance}
        \subsubsection{Subgroup analysis}
            \begin{figure}[h]
              \centering
              \fbox{\includegraphics[width=0.95\textwidth]{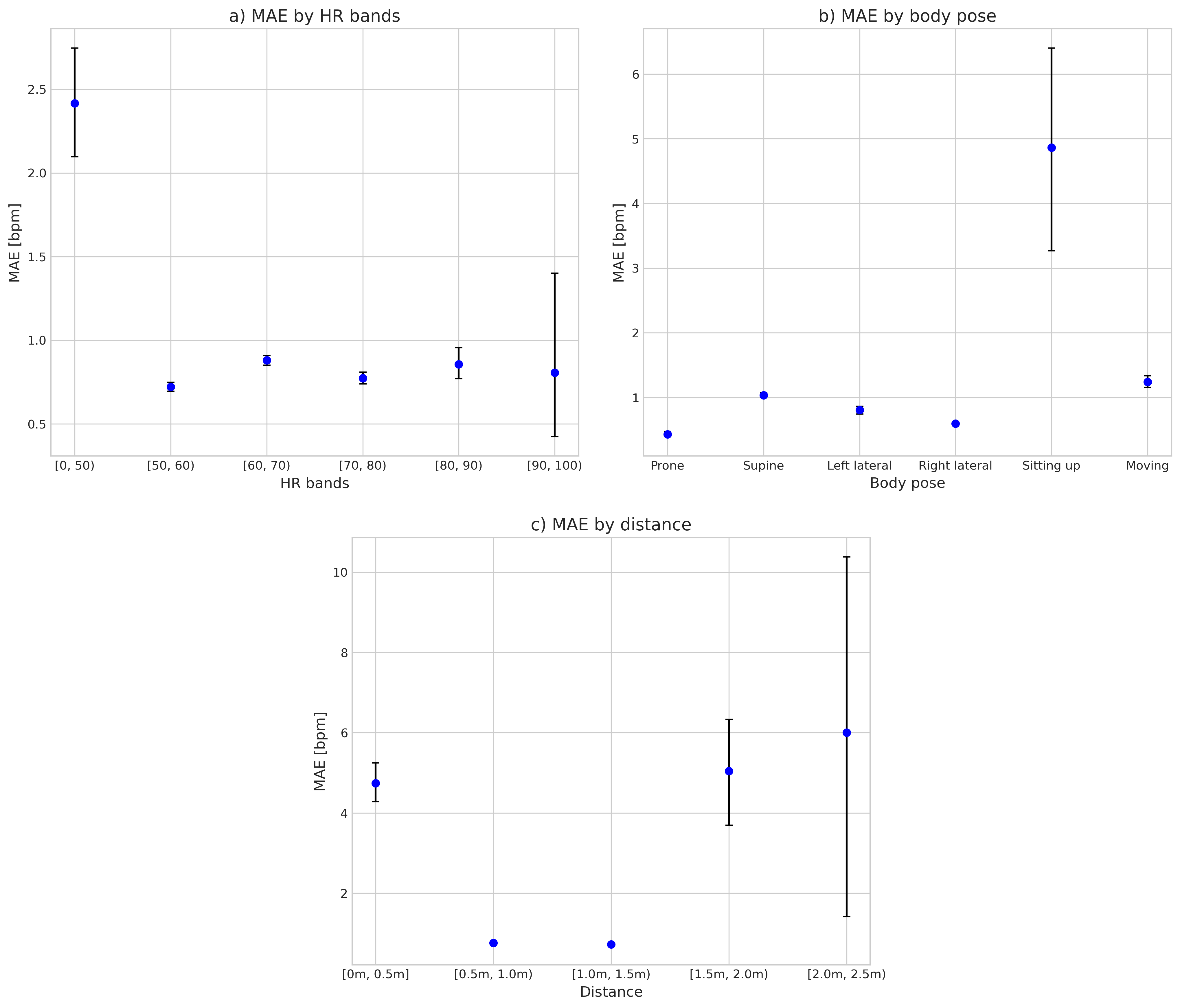} }
              \caption{Model performance on mm-wave FMCW radar data including 95\% confidence interval by a) HR bands, b) body position, c) user distance from the radar.}
              \label{fig:figs1}
            \end{figure}
        \subsubsection{Ablations}
        \label{sec:FMCW_ablations}
        We considered a range of ablations to verify that the model architecture as well as the features selections leads to the best performance.
        The ablation metrics are reported on the validation set.

            \paragraph{Model architectures}
            \label{sec:FMCW_model_architecture}
            We compare the performance of our proposed model architecture with the following standard architectures:
            \begin{enumerate}[label=\alph*)]
                \item \textbf{2D ResNet18:} 2D ResNet18, followed by a fully connected layer.
                \item \textbf{2D ResNet (2, 2, 2, 2, 2, 2):} 2D ResNet which consists of 6 stages, with 2 ResNet blocks each, to match the number of stages in our proposed 2D~+~1D ResNet architecture, followed by a fully connected layer.
                \item \textbf{1D ResNet18 + max pooling:} 1D ResNet18 per feature followed by max pooling across features dimensions, followed by a fully connected layer.
                \item \textbf{1D ResNet (2, 2, 2, 2, 2, 2) + max pooling:} 1D ResNet which consists of 6 stages, with 2 ResNet blocks each, per feature, followed by max pooling across features dimensions, followed by a fully connected layer.
            \end{enumerate}
            Our proposed 2D~+~1D ResNet architecture outperforms standalone 2D or 1D ResNet by a large margin with a 0.89 bpm validation set MAE versus an 1.30 bpm MAE of when using a standard 1D ResNet per feature (see Table~\ref{tab:tabs1}).    
            \begin{table}[h]
             \caption{Comparison of the performance of our proposed 2D~+~1D ResNet model architectures and other standard architectures.}
              \centering
              \begin{tabular}{l c}
                \toprule
                \multicolumn{2}{c}{\textbf{32 range bins}}  \\
                \multicolumn{2}{c}{\textbf{3 antennas concatenated}}  \\
                \multicolumn{2}{c}{\textbf{Unwrapped angle + magnitude}}  \\
                \midrule
                \textbf{2D~+~1D ResNet (ours)}                           & \textbf{0.89}       \\
                \midrule
                2D ResNet18                                    & 1.80       \\
                2D ResNet (2, 2, 2, 2, 2, 2)                   & 1.42       \\
                1D ResNet18 + max pooling                     & 1.62       \\
                1D ResNet (2, 2, 2, 2, 2, 2) + max pooling    & 1.30       \\
                \bottomrule
                \end{tabular}
              \label{tab:tabs1}
            \end{table}
            \paragraph{Feature selection.}
            \label{sec:FMCW_feature_selection}
            We performed an ablation on including: only unwrapped angle, only magnitude, or both those features for each range bin, as well as considering various antenna combinations, namely:
            \begin{enumerate}[label=\alph*)]
            \item \textbf{Split antennas:} Split each antenna into a separate example, so that the model is trained / evaluated on a single antenna at the time.
            \item \textbf{3 antennas concatenated:} Concatenate all 3 antennas into a single example, so that the model always considers data from all antennas.
            \item \textbf{Single antenna:} Only use a single antenna to train and evaluate the model.
            \item \textbf{Two antennas concatenated:} Use a pair of two antennas to train and evaluate the model.
            \item \textbf{PCA beamforming:} Perform a PCA beamforming to combine the data from 3 or 2 antennas before performing feature extraction.
            \end{enumerate}
            The results of this ablation are presented in Table~\ref{tab:tabs2} and Table~\ref{tab:tabs3}. It clearly indicates that including both features: unwrapped angle and magnitude, reduced MAE by up to 65\%. It also shows concatenating multiple antennas improved the MAE. More concretely, in the case of FMCW radar used in this study, concatenating only 2 antennas (antenna 1 and 2) leads to comparable performance as using all 3 antennas; this is due to the fact, that in this specific radar, the performance of antenna \#0 was very poor.            
            \begin{table}[h]
             \caption{Ablations on feature selection and antenna combinations when selecting 64 range bins.}
              \centering
              \begin{tabular}{lccc}
                \toprule
                \multicolumn{4}{c}{\textbf{2D~+~1D ResNet}} \\
                \multicolumn{4}{c}{\textbf{64 range bins}} \\
                \midrule
                 & \textbf{Unwrapped angle} & \textbf{Magnitude} & \textbf{Unwrapped angle + Magnitude} \\
                \midrule
                Split antennas              & 3.77                     & 2.99               & 1.29                                 \\
                3 antennas concatenated     & 1.52                     & 1.41               & \textbf{0.89}                                 \\
                Antenna \#0                 & 4.78                     & 4.73               & 2.47                                 \\
                Antenna \#1                 & 4.39                     & 3.96               & 2.18                                 \\
                Antenna \#2                 & 2.47                     & 2.21               & 1.16                                 \\
                \bottomrule
              \end{tabular}
              \label{tab:tabs2}
            \end{table}
            \begin{table}[h]
             \caption{Ablations on antenna combinations when using both: unwrapped angle and magnitude features and selecting 64 range bins.}
              \centering
              \begin{tabular}{lc}
                \toprule
                    \multicolumn{2}{c}{\textbf{2D~+~1D ResNet}} \\
                    \multicolumn{2}{c}{\textbf{64 range bins}} \\
                    \multicolumn{2}{c}{\textbf{Unwrapped angle + magnitude}} \\
                    
                    \midrule
                    2D~+~1D ResNet (ours)                      & 1.29           \\
                    3 antennas concatenated                  & \textbf{0.89}           \\
                    3 antennas pca beamforming               & 1.23           \\
                    2 antennas pca beamforming \#1\#2          & 1.26           \\
                    Antenna \#0                              & 2.47           \\
                    Antenna \#1                              & 2.18           \\
                    Antenna \#2                              & 1.16           \\
                    Antenna \#0\#1 concatenated                & 1.34           \\
                    Antenna \#0\#2 concatenated                & 1.13           \\
                    Antenna \#1\#2 concatenated                & \textbf{0.87}           \\
                    \bottomrule
              \end{tabular}
              \label{tab:tabs3}
            \end{table}
            \paragraph{Number of range bins selection.}
            \label{sec:FMCW_range_bins}
            We also performed ablations on the number of range bins to select. As shown in the Table~\ref{tab:tabs4}, the performance of 0.89 MAE is achieved when including 64 or 32 range bins, but drops significantly to 1.32 MAE when only including 16 range bins. 
            \begin{table}[h]
             \caption{Ablations on number of range bins selected when using 3 antennas concatenated and both: unwrapped angle and magnitude features.}
              \centering
              \begin{tabular}{lc}
                \toprule
                \multicolumn{2}{c}{\textbf{2D~+~1D ResNet}} \\
                \multicolumn{2}{c}{\textbf{3 antennas concatenated}} \\
                \multicolumn{2}{c}{\textbf{Unwrapped angle + magnitude}} \\
                \midrule
                64 range bins & 0.89       \\
                32 range bins & \textbf{0.89}       \\
                16 range bins & 1.32       \\
                \bottomrule
                \end{tabular}
              \label{tab:tabs4}
            \end{table}

    \subsection{IR-UWB radar performance}
        \subsubsection{Subgroup analysis}
            \begin{figure}[H]
              \centering
              \fbox{\includegraphics[width=0.95\textwidth]{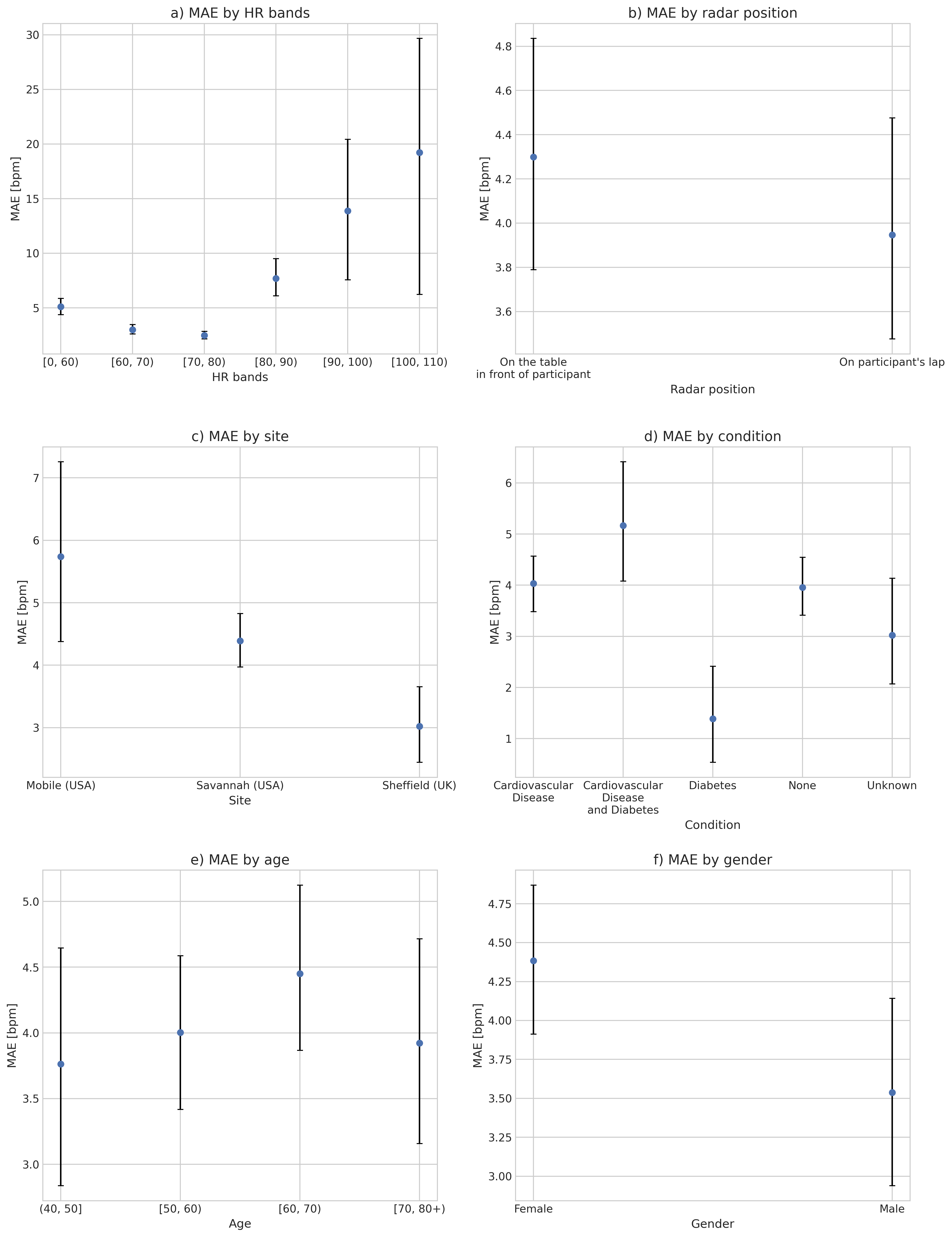} }
              \caption{Model performance on IR-UWB radar data including 95\% confidence interval by a) HR bands, b) radar position, c) site, d) condition, e) age and f) gender.}
              \label{fig:figs2}
            \end{figure}
        \subsubsection{Ablations}
        The ablation metrics are reported on the validation set.
            \paragraph{Baseline model ablations}
            When training the baseline model on IR-UWB dataset from scratch, we performed similar ablations to those described in ~\ref{sec:FMCW_feature_selection} and ~\ref{sec:FMCW_range_bins}.            
            \subparagraph{Baseline model feature selection.}
            We performed an ablation on including: only unwrapped angle, only magnitude, or both those features for each range bin, as well as considering various antenna combinations, namely:
            \begin{enumerate}[label=\alph*)]
                \item \textbf{Split antennas:} Split each antenna into a separate example, so that the model is trained / evaluated on a single antenna at the time.
                \item \textbf{2 antennas concatenated:} Concatenate all 2 antennas into a single example, so that the model always considers data from all antennas.
                \item \textbf{Single antenna:} Only use a single antenna to train and evaluate the model.
            \end{enumerate}
            The results of ablations are presented in the Table~\ref{tab:tabs5}. It indicates that to train a baseline model, splitting each antenna into separate examples gives the best performance. It also shows that including both unwrapped angle and magnitude features improves the performance of the model slightly, but only when training on a single antenna.
            \begin{table}[h]
             \caption{Ablations on feature selection and antenna combinations when selecting 6 range bins.}
              \centering
              \begin{tabular}{lccc}
                \toprule
                \multicolumn{4}{c}{\textbf{2D~+~1D ResNet}} \\
                \multicolumn{4}{c}{\textbf{6 range bins}} \\
                \midrule
                & \textbf{Unwrapped angle} & \textbf{Magnitude} & \textbf{Unwrapped angle + Magnitude} \\
                \midrule
                Split antennas          & 5.9                      & 6.2                & 5.9                                  \\
                2 antennas concatenated & 8.1                      & 8.1                & 8.3                                  \\
                Antenna \#0             & 6.8                      & 6.3                & 6.2                                  \\
                Antenna \#1             & 7.2                      & 7.5                & 7.0                                  \\
                \bottomrule
                \end{tabular}
              \label{tab:tabs5}
            \end{table}
            \subparagraph{Baseline model number of range bins selection.}
            \label{sec:UWB_baseline_range_bins}
            We also performed ablations on the number of range bins to select. As shown in the Table~\ref{tab:tabs6}, the performance of the model is best when only using a single range bin.
            \begin{table}[h]
             \caption{Ablations on number of range bins selected when using 2 antennas split into separate examples and both: unwrapped angle and magnitude features.}
              \centering
              \begin{tabular}{lc}
                \toprule
                \multicolumn{2}{c}{\textbf{2D~+~1D ResNet}} \\
                \multicolumn{2}{c}{\textbf{2 antennas split into separate examples}} \\
                \multicolumn{2}{c}{\textbf{Unwrapped angle + magnitude}} \\
                \midrule
                6 range bins & 5.9 \\
                4 range bins & 6.2 \\
                3 range bins & 5.7 \\
                1 range bin  & \textbf{5.4} \\
                \bottomrule
                \end{tabular}
              \label{tab:tabs6}
            \end{table}
        \paragraph{Transfer learning ablations}
        The additional preprocessing steps aimed to address the following issues: different number of receiving antennas and different range resolution of two radar systems.
        The goal was to maintain the same dimensionality of the data from two radar systems, despite their different characteristics.
        
        To address a different number of antennas, we evaluated the following techniques:

        \begin{enumerate}[label=\alph*)]
            \item Choose a single antenna when training and fine-tuning. 
            \item Choose a pair of two antennas when training; fine-tune on all two available antennas.
            \item Perform a PCA beamforming prior to feature extraction step; train and fine-tune on features extracted from one beamformed antenna.
            \item Perform a convolution step prior to 2D~+~1D ResNet to combine features from 3 antennas during training, and from 2 antennas during fine-tuning.
        \end{enumerate}
        
        To address a different range resolution we evaluated the following techniques:
        \begin{enumerate}[label=\alph*)]
            \item Downsample the range bins of the mm-wave FMCW dataset by averaging 10 range bins into 1 range bin when training the model; this is conceptually equivalent to downsampling an image. The advantage of this method is that it results in the same physical area being considered on each dataset.
            \item Select a single range bin when training the model and single range bin when fine-tuning the model. According to ablations ~\ref{sec:UWB_baseline_range_bins}, using a single range bin from IR-UWB radar gives the best performance. This attempts to address potential drawbacks of the method a), that averaging waveforms may not be ideal if two waveforms represent the same frequencies but have the opposite phase. The downside is that this method results in a different physical area being considered on each dataset. 
        \end{enumerate}
        The results of the ablations are presented in the Table~\ref{tab:tabs7} and Table~\ref{tab:tabs8}. We show, that training the model on either antenna \#1 or \#2 from the mm-wave FMCW radar dataset and fine-tuning it on the antenna \#0 gives the best performance. We also show that selecting a single range bin when pre-training and fine-tuning the model gives best results.
            \begin{table}[H]
             \caption{Ablations on antenna configurations when the base model was trained on 60 range bins downsampled to 6 range bins.}
              \centering
              \begin{tabular}{p{3cm} p{4.5cm} p{3cm} p{4.5cm}}
                \toprule
                \multicolumn{4}{c}{\textbf{2D~+~1D ResNet}} \\
                \multicolumn{4}{c}{\textbf{Base model trained on 60 range bins of mm-wave FMCW radar, which were averaged to 6 range bins}} \\
                \multicolumn{4}{c}{\textbf{Final model fine tuned on 6 range bins of IR-UWB radar}} \\
                \midrule
                \textbf{Base model\newline antenna\newline configuration} & \textbf{MAE of base model\newline on FMCW radar dataset} & \textbf{Final model\newline antenna\newline configuration} & \textbf{MAE of fine tuned model\newline on IR-UWB radar database} \\
                \midrule
                Antenna \#0 & 6.35 & Antenna \#0 & 4.3 \\
                 &  & Antenna \#1 & 5.5 \\
                 &  & Split antennas & 4.5 \\
                 \midrule
                Antenna \#1 & 5.81 & Antenna \#0 & 4.0 \\
                 &  & Antenna \#1 & 4.9 \\
                 &  & Split antennas & 4.7 \\
                 \midrule
                Antenna \#2 & 3.94 & Antenna \#0 & 4.0 \\
                 &  & Antenna \#1 & 5.3 \\
                 &  & Split antennas & 4.2 \\
                  \midrule
                Antenna \#1 \#2 & 3.10 & Antennas \#0 \#1 & 5.8 \\
                 &  & Antennas \#1 \#0 & 5.7 \\
                  \midrule
                3 antennas PCA beamforming & 3.97 & 2 antennas PCA beamforming & 4.4 \\
                \bottomrule
                \end{tabular}
              \label{tab:tabs7}
            \end{table}
            \begin{table}[H]
             \caption{Ablations on antenna configurations when the base model was trained on a single range bin.}
              \centering
              \begin{tabular}{p{3cm} p{4.5cm} p{3cm} p{4.5cm}}
                \toprule
                \multicolumn{4}{c}{\textbf{2D~+~1D ResNet}} \\
                \multicolumn{4}{c}{\textbf{Base model trained on 1 range bin of mm-wave FMCW radar}} \\
                \multicolumn{4}{c}{\textbf{Final model fine tuned on 1 range bin of IR-UWB radar}} \\
                \midrule
                \textbf{Base model\newline antenna\newline configuration} & \textbf{MAE of base model\newline on FMCW radar dataset} & \textbf{Final model\newline antenna\newline configuration} & \textbf{MAE of fine tuned model\newline on IR-UWB radar database} \\
                \midrule
                Antenna \#2 & 2.27 & Antenna \#0 & 3.8 \\
                Antenna \#1 \#2 & 2.50 & Antenna \#1 \#0 & 5.5 \\
                3 antennas PCA beamforming & 2.80 & 2 antennas PCA beamforming & 5.2 \\
                3 antennas Conv beamforming & 2.20 & 2 antennas Conv beamforming
 & 4.7 \\
                \bottomrule
                \end{tabular}
              \label{tab:tabs8}
            \end{table}
        \paragraph{Transfer learning data augmentation ablations.}
        \label{sec:UWB_augmentations}
        We also applied the following data augmentation techniques when training the model on mm-wave FMCW and fine-tuning it on IR-UWB:
        \begin{enumerate}[label=\alph*)]
            \item Apply Gaussian noise with zero mean and standard deviation of 0.0005 to the input data during training and fine tuning stage.
            \item Swap the order of features (magnitude and unwrapped angle) at random for each range bin per antenna when training on mm-wave FMCW dataset; this is to make the model more robust to the order of features and their significance.
            \item Swap the order of features (magnitude and unwrapped angle) for each range bin per antenna during the fine tuning stage; the rationale behind this is that those two features have a different significance depending on the radar type.
        \end{enumerate}
        The results of the ablations are presented in Table~\ref{tab:tabs9} and Table~\ref{tab:tabs10}. We show that applying Gausian noise when pre-training and fine-tuning the model improves the model performance. We also show that swapping the order of features when fine-tuning the model improves the model's performance.

\begin{figure}[h]
              \centering
              \fbox{\includegraphics[width=0.8\textwidth]{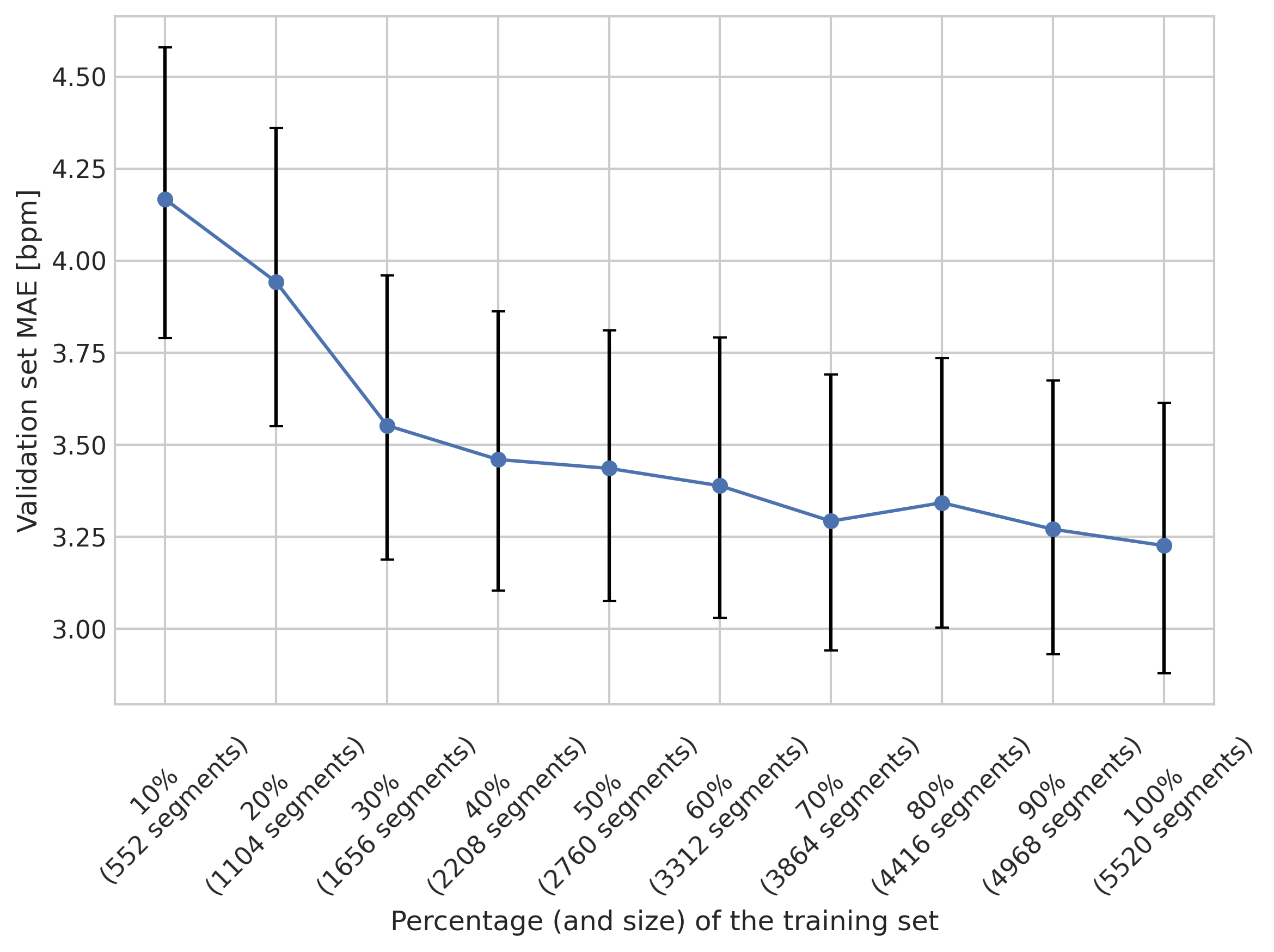} }
              \caption{The effect of size of the training set on the model performance on the validation set.}
              \label{fig:figs3}
            \end{figure}

            \begin{table}[H]
             \caption{Ablations on data augmentations when training the base model on mm-wave FMCW dataset.}
              \centering
              \begin{tabular}{p{6cm} p{4.5cm} p{4.5cm}}
                \toprule
                \multicolumn{3}{c}{\textbf{2D~+~1D ResNet}} \\
                \multicolumn{3}{c}{\textbf{Base model trained on antenna \#2 and 1 range bin of mm-wave FMCW radar}} \\
                \multicolumn{3}{c}{\textbf{Final model fine tuned on antenna \#0 and 1 range bin of IR-UWB radar}} \\
                \midrule
                \textbf{Base model\newline augmentations} & \textbf{MAE of base model\newline on FMCW radar dataset} & \textbf{MAE of fine tuned model\newline on IR-UWB radar database} \\
                \midrule
                No augmentations & 2.27 & 3.8 \\
                Gaussian noise with probability=0.7 & 2.45 & \textbf{3.7} \\
                Randomly flip order of features + Gaussian noise with probability=0.6 & 2.73 & 4.2 \\
                \bottomrule
                \end{tabular}
              \label{tab:tabs9}
            \end{table}
            \begin{table}[H]
             \caption{Adding data augmentations during fine tuning on IR-UWB dataset.}
              \centering
              \begin{tabular}{p{4.5cm} p{6cm} p{4.5cm}}
                \toprule
                \multicolumn{3}{c}{\textbf{2D~+~1D ResNet}} \\
                \multicolumn{3}{c}{\textbf{Base model trained on antenna \#2 and 1 range bin of mm-wave FMCW radar with Gausian noise with probability=0.7}} \\
                \multicolumn{3}{c}{\textbf{Final model fine tuned on antenna \#0 and 1 range bin of IR-UWB radar}} \\
                \midrule
                \textbf{MAE of base model\newline on FMCW radar dataset} & \textbf{Final model \newline augmentations} & \textbf{MAE of fine tuned model\newline on IR-UWB radar database} \\
                \midrule
                2.45 & No augmentations & 3.7 \\
                 & Flip order of features & 3.4 \\
                 & Flip order of features + Gaussian noise with probability=0.6 & \textbf{3.2} \\
                \bottomrule
                \end{tabular}
              \label{tab:tabs10}
            \end{table}
        \paragraph{Transfer learning size of training dataset ablations.}
        \label{sec:UWB_train_set_size}
        The ablation presented in the Figure~\ref{fig:figs3} and Table~\ref{tab:tabs11} show that using the first 40\% of the training set during the fine-tuning stage brings the most significant model improvements, beyond which point the gain is much smaller, linear to the dataset size. This suggests that collecting more data would continue to give small performance gains.
            \begin{table}[h]
             \caption{The effect of size of the training set on the model performance on the evaluation set.}
              \centering
              \begin{tabular}{cc}
                \toprule
                \textbf{\% of train set} & \textbf{Validation MAE [95\% CI]} \\
                \midrule
                10\%                   & 4.2 [3.8, 4.6]                  \\
                20\%                   & 3.9 [3.6, 4.4]                  \\
                30\%                   & 3.6 [3.2, 4.0]                  \\
                40\%                   & 3.5 [3.1, 3.9]                  \\
                50\%                   & 3.4 [3.1, 3.8]                  \\
                60\%                   & 3.4 [3.0, 3.8]                  \\
                70\%                   & 3.3 [2.9, 3.7]                  \\
                80\%                   & 3.3 [3.0, 3.7]                  \\
                90\%                   & 3.3 [2.9, 3.7]                  \\
                100\%                  & 3.2 [2.9, 3.6]                  \\
                \bottomrule
                \end{tabular}
              \label{tab:tabs11}
            \end{table}
        \paragraph{Transfer learning model architecture ablations.}
        \label{sec:UWB_model_architecture}
        Based on all the ablations presented above, the best performing model only used 2 features as an input: unwrapped angle and magnitude from a single antenna and a single range bin. We therefore accessed, if using our proposed 2D~+~1D ResNet model architecture still brings benefits compared to other, more standard model architectures. We performed the following comparison experiments:
        \begin{enumerate}[label=\alph*)]
            \item \textbf{1D ResNet (2,2,2,2,2,2) + max pooling:} 1D ResNet which consists of 6 stages, with 2 ResNet blocks each and the feature width factor of 2, per feature, followed by max pooling across features dimensions, followed by a fully connected layer.
            \item \textbf{1D ResNet18 + max pooling:} 1D ResNet18 per feature with the feature width factor of 2, followed by max pooling across features dimensions, followed by a fully connected layer.
            \item \textbf{Single 2D convolution + 1D ResNet (2,2,2,2,2,2):} Single 2D convolution followed by 1D ResNet which consists of 6 stages, with 2 ResNet blocks each and the feature width factor of 2, followed by a fully connected layer.
        \end{enumerate}
        As shown in the Table~\ref{tab:tabs12}, the 1D ResNet model with 6 stages per feature, followed by max pooling across features dimensions achieves the same performance of 3.8 MAE to our proposed model architecture, however it is a much bigger model, requiring over 15x more parameters. 1D ResNet with the comparable number of parameters achieves much worse performance of 4.8 MAE compared to our proposed architecture.  
            \begin{table}[h]
             \caption{Comparing our proposed 2D~+~1D ResNet model architectures with other standard architectures.}
              \centering
              \begin{tabular}{p{4.5cm} p{3cm} p{3.5cm} p{3.5cm}}
                \toprule
                \textbf{Model architecture}                            & \textbf{Number of model's parameters} & \textbf{MAE of base model on FMCW radar dataset} & \textbf{MAE of fine tuned model on IR-UWB radar database} \\
                \midrule
                \textbf{2D~+~1D ResNet (ours)}                              & \textbf{16M}                                   & \textbf{2.27}                                          & \textbf{3.8}                                                     \\
                \midrule
                1D ResNet (2,2,2,2,2,2) + max pooling            & 250M                                  & 2.77                                           & 3.8                                                     \\
                1D ResNet18 + max pooling                         & 15M                                   & 4.70                                          & 4.8                                                     \\
                Single 2D convolution + 1D ResNet (2,2,2,2,2,2) & 250M                                  & 3.70                                          & 4.0                                                     \\
                \bottomrule
                \end{tabular}
              \label{tab:tabs12}
            \end{table}
        %
    \subsubsection{Improving model performance in high HR subgroups via data augmentation.}
    \label{sec:UWB_high_hr}
\begin{figure}
          \centering
          \fbox{\includegraphics[width=0.95\textwidth]{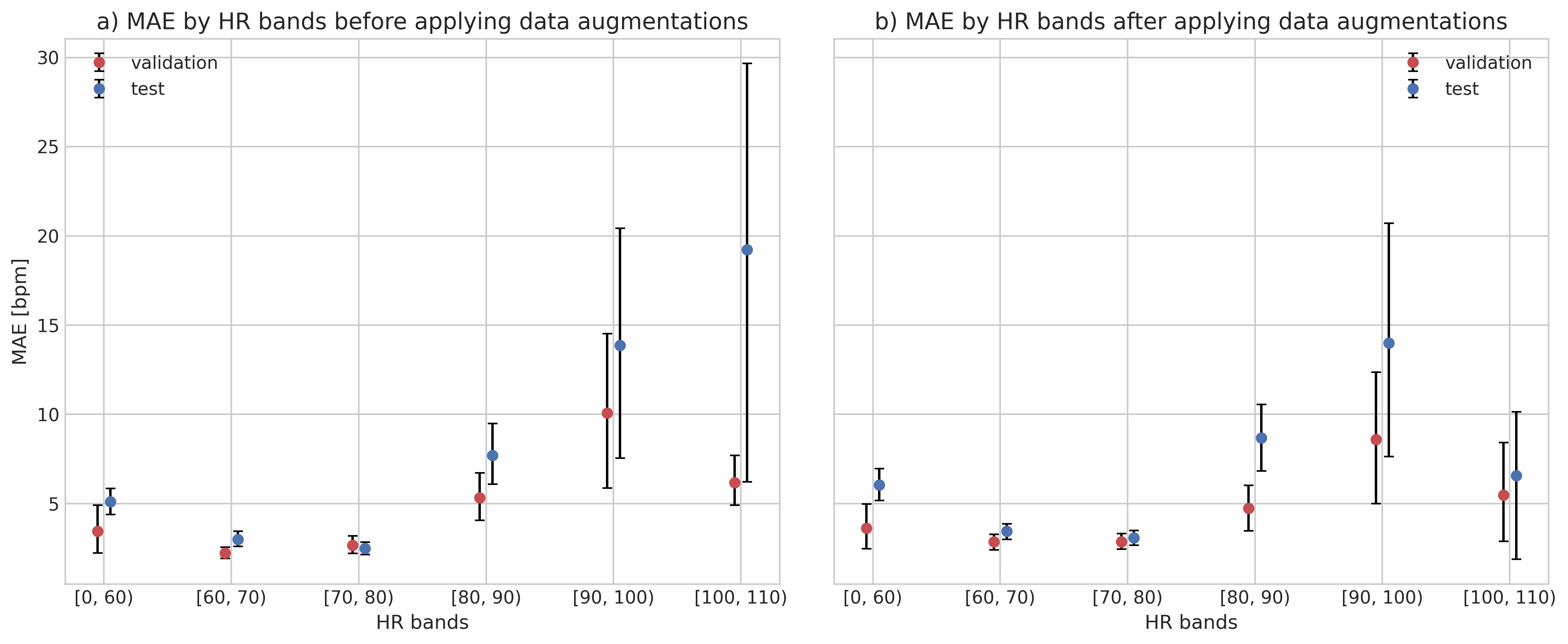} }
          \caption{Model performance on validation and test sets of IR-UWB radar data including 95\% confidence interval by HR bands a) before and b) after applying data augmentations.}
          \label{fig:figs4}
        \end{figure}

        We did preliminary experiments to mitigate high MAE this issue by applying additional data augmentation when fine-tuning the model on IR-UWB, to accelerate HR in samples with the HR equal or greater than 70 bpm by a multiplier between 1.0 and 1.2, at a probability of 0.5. We also upweighted the samples with the HR of 90 bpm and higher by the factor of 1.5. While this decreased the overall MAE from 3.2 bpm to 3.4 bpm on the validation set and from 4.1 bpm to 4.8 bpm on the test set, it improved the performance for the higher HR subgroups (see Table~\ref{tab:tabs13} and Table~\ref{tab:tabs14} and Figure~\ref{fig:figs4}).
        \begin{table}[htb]
         \caption{Model performance on validation set of IR-UWB radar data including 95\% confidence interval by HR bands before and after applying data augmentations.}
          \centering
          \begin{tabular}{p{2cm} p{4.5cm} p{4.5cm} p{2cm}}
            \toprule
            & \multicolumn{2}{c}{\textbf{Validation MAE [95\% CI]}} & \\
            \textbf{HR band} & \textbf{No augmentation} & \textbf{Accelerate HR > 70 bpm \newline Upweight HR > 90bpm} & \textbf{\% of validation samples} \\
            \midrule
            Overall          & 3.2 [2.9, 3.6]                    & 3.4 [3.1, 3.8]                    & $-$                               \\
            \midrule
            \text{[}0, 60)         & 3.5 [2.2, 4.9]                    & 3.6 [2.5, 5.0]                    & 12.3                              \\
            \text{[}60, 70)        & 2.2 [1.9, 2.6]                    & 2.9 [2.4, 3.3]                    & 43.5                              \\
            \text{[}70, 80)        & 2.7 [2.2, 3.2]                    & 2.9 [2.5, 3.3]                    & 26.1                              \\
            \text{[}80, 90)        & 5.3 [4.1, 6.7]                    & 4.7 [3.5, 6.0]                    & 13.3                              \\
            \text{[}90, 100)       & 10.1 [5.9, 14.5]                  & 8.6 [5.0, 12.4]                   & 3.2                               \\
            \text{[}100, 110)      & 6.2 [4.9, 7.7]                    & 5.5 [2.9, 8.4]                    & 1.5                               \\
            \bottomrule
            \end{tabular}
          \label{tab:tabs13}
        \end{table}
        \begin{table}[htb]
         \caption{Model performance on test set of IR-UWB radar data including 95\% confidence interval by HR bands before and after applying data augmentations.}
          \centering
          \begin{tabular}{p{2cm} p{4.5cm} p{4.5cm} p{2cm}}
            \toprule
            & \multicolumn{2}{c}{\textbf{Test MAE [95\% CI]}} & \\
            \textbf{HR band} & \textbf{No augmentation} & \textbf{Accelerate HR > 70 bpm \newline Upweight HR > 90bpm} & \textbf{\% of validation samples} \\
            \midrule
            Overall          & 4.1 [3.8, 4.5]            & 4.7 [4.3, 5.1]            & $-$                         \\
            \text{[}0, 60)         & 5.1 [4.4, 5.9]            & 6.1 [5.2, 7.0]            & 28.4                        \\
            \text{[}60, 70)        & 3.0 [2.6, 3.5]            & 3.5 [3.0, 3.9]            & 32.2                        \\
            \text{[}70, 80)        & 2.5 [2.1, 2.8]            & 3.1 [2.7, 3.5]            & 28.2                        \\
            \text{[}80, 90)        & 7.7 [6.1, 9.5]            & 8.7 [6.8, 10.6]           & 9.2                         \\
            \text{[}90, 100)       & 13.9 [7.6, 20.4]          & 14.0 [7.6, 20.7]          & 1.6                         \\
            \text{[}100, 110)      & 19.2 [6.2, 29.7]          & 6.6 [1.9, 10.1]           & 0.4                         \\
            \bottomrule
            \end{tabular}
          \label{tab:tabs14}
        \end{table}

    \end{appendices}

\end{document}